\documentclass[fleqn,10pt]{wlscirep}

\usepackage{graphicx}
\usepackage{subcaption}
\usepackage{mwe}
\usepackage{soul}

\title{Quantifying the Global Support Network for Non-State Armed Groups (NAGs)}

\author[1,2,*]{Weiran Cai}
\author[3]{Belgin San-Akca}
\author[2,4,5]{Jordan Snyder}
\author[1]{Grayson Gordon}
\author[6]{Zeev Maoz}
\author[1,2,7,8]{Raissa M. D'Souza}
\affil[1]{Department of Computer Science, University of California at Davis, 1 Shields Ave., Davis, CA 95616, USA}
\affil[2]{Complexity Sciences Center, UC Davis, CA 95616, USA}
\affil[3]{Department of International Relations, Ko\c{c} University, Istanbul 34450, Turkey}
\affil[4]{Department of Mathematics, UC Davis, CA 95616, USA}
\affil[5]{Department of Applied Mathematics, University of Washington, Seattle, WA 98103, USA}
\affil[6]{Department of Political Sciences, UC Davis, 1 Shields Ave., Davis, CA 95616, USA}
\affil[7]{Department of Mechanical and Aerospace Engineering, UC Davis, 1 Shields Ave., Davis, CA 95616, USA}
\affil[8]{Santa Fe Institute, 1399 Hyde Park Rd, Santa Fe, NM 87501}
\affil[*]{wrcai@ucdavis.edu}

\begin{abstract}

Human history has been shaped by armed conflicts. Rather than large-scale interstate wars, low-intensity attacks have been more prevalent in the post-World War era. These attacks are often carried out by non-state armed groups (NAGs), which are supported by host states (HSs). We analyze the global bipartite network of NAG-HS support and its evolution over the period of 1945-2010. We find striking parallels to ecological networks such as mutualistic and parasitic forms of support, and a nested and modular network architecture. The nestedness emerges from preferential behaviors: highly connected players are more likely to both gain and lose connections. Long-persisting major modules are identified, reflecting both regional and trans-regional interests, which show significant turnover in their membership, contrary to the transitory ones. Revealing this architecture further enables the identification of actor’s roles and provide insights for effective intervention strategies.
\end{abstract}
\begin{document}

\flushbottom
\maketitle
%
%
\thispagestyle{empty}

\section{Introduction}

Military confrontation is one of the key factors that have shaped human history. While instances of large-scale interstate warfare have decreased in the post-World War era, internal wars and low-intensity conflicts carried out by non-state armed groups (NAGs) against nation-states have become increasingly common \cite{San-Akca:2016,Maoz:2012,Horowitz:2014,Phillips:2018,LaFree:2009,Phillips:2015,Freilicha:2015,Pinker:2012,Gleditsch:2013,Gleditsch:2016,Byman:2001,Kalyvas:2010}. NAGs include rebel, insurgent, guerrilla, or terrorist groups that engage in violent activities targeting the government, citizens, or institutions of nation-states. Many of these NAGs are cultivated by a complex network of supporting host states (HSs) --- states external to the locus of the internal 
conflict between the NAG and the target government. Examples include U.S. support to the Contras in Nicaragua in the 1980s, Israeli and Iranian support for Kurdish rebels in Iraq during the 1960s and 1970s, and NATO support for the Libyan rebels in 2011 \cite{San-Akca:2016}. The 
HSs support NAGs by providing them with military and economic aid, sanctuary to leaders or members of NAGs, logistics, or training.
The bipartite relation between NAGs and host states has become a common mode of foreign policy conduct \cite{San-Akca:2016,Maoz:2012,Gleditsch:2016,Byman:2001,Kalyvas:2010}, yet we lack reliable real-time data on NAG-HS relations as most of these interactions are covert; public information usually emerges \emph{ex-post facto.} The few existing studies have focused on specific NAG-HS relations \cite{LaFree:2009,Freilicha:2015,Popovic:2017,Salehyan:2010,Gleditsch:2008}. 
The entire ecosystem of NAG and HS interactions and its evolution over time remains an uncharted territory.
Revealing the underlying patterns and mechanisms behind the NAG-HS support network is important for analytical insights and policy implications.

We construct the global network of bipartite support between NAGs and HSs involving all military conflicts over the period of 1945-2010 and measure its global topological characteristics. Two types of links are present in the network: A HS can provide a NAG with either intentional (``active'') or de facto (``passive") support. Active support entails a deliberate decision by a HS to form an alliance with a NAG in order to further common interests, such as the Israeli and Iranian support for the Kurds in the 1960s and 1970s \cite{maoz2006}. Passive support refers to situations where a NAG operates or extracts resources from a HS, while the HS does not engage in a deliberate action or create channels to provide any support. 
Examples of passive support include the IRA operations in the Irish community in the US during the 1980s and 1990s \cite{guelke1996}, or the fundraising for Palestinian insurgents among Islamic communities in the US prior to September 11, 2001 \cite{levitt2008}. 
Likewise, a NAG may operate in a HS whose institutions are weak and incapable of resisting the NAG's activities (e.g., the PLO operation in Lebanon over the 1970-1982 period) \cite{San-Akca:2016, San-Akca:2014}. 

%
By tracking the changes in the players and supporting relations, we find that the network evolves via a mechanism of generalized preferential attachment and detachment: highly connected players are more likely to both gain and lose connections relative to peripheral players. Such evolution leads to characteristic patterns typically associated with ecological networks, such as plant-pollinator networks, seed-frugivore networks \cite{Bascompte:2003,Olesen:2007} and 
host-parasite networks \cite{Poulin:2010,Graham:2009}, which similarly involve tradeoffs through cooperative or exploitative relations.  
%
%
In particular, we find a robust nested and modular architecture over a broad time span. Nestedness entails a nonrandom pattern of generalist-specialist structure, which is a crucial factor determining the structural stability of a network \cite{Okuyama:2008,Bastolla:2009,Rohr:2014}. Modularity indicates the extent to which the network is organized into clearly delimited clusters. Such patterns are also frequently observed in socio-economic systems, from the network of bankers \cite{May:2008} to the designer-manufacturer network of New York garment industry \cite{Saavedra:2009}. 

Using a temporal community detection algorithm we find that there are nine major modules that persist over time and many small, short-lived modules. The persistent modules reveal both regional and trans-regional composition and historical connections. These major modules display significant variation in their membership, whereas the transitory ones show high consistency. Furthermore, we identify the roles of nodes and find that if a NAG or HS is involved in both the active and passive support subnetworks they play the similar role of hub or peripehral in both subnetworks. The discussion section addresses some of the implications of these findings. We suggest that our understanding of robustness and interventions in ecological bipartite networks may be useful for developing efficient disruptions of malign relations in the NAG-HS network, where we can treat the support as mutualistic and parasitic relationships.  


\section{Results}
\begin{figure}[t]
\hskip 1.8cm
\includegraphics[width=0.75\textwidth]{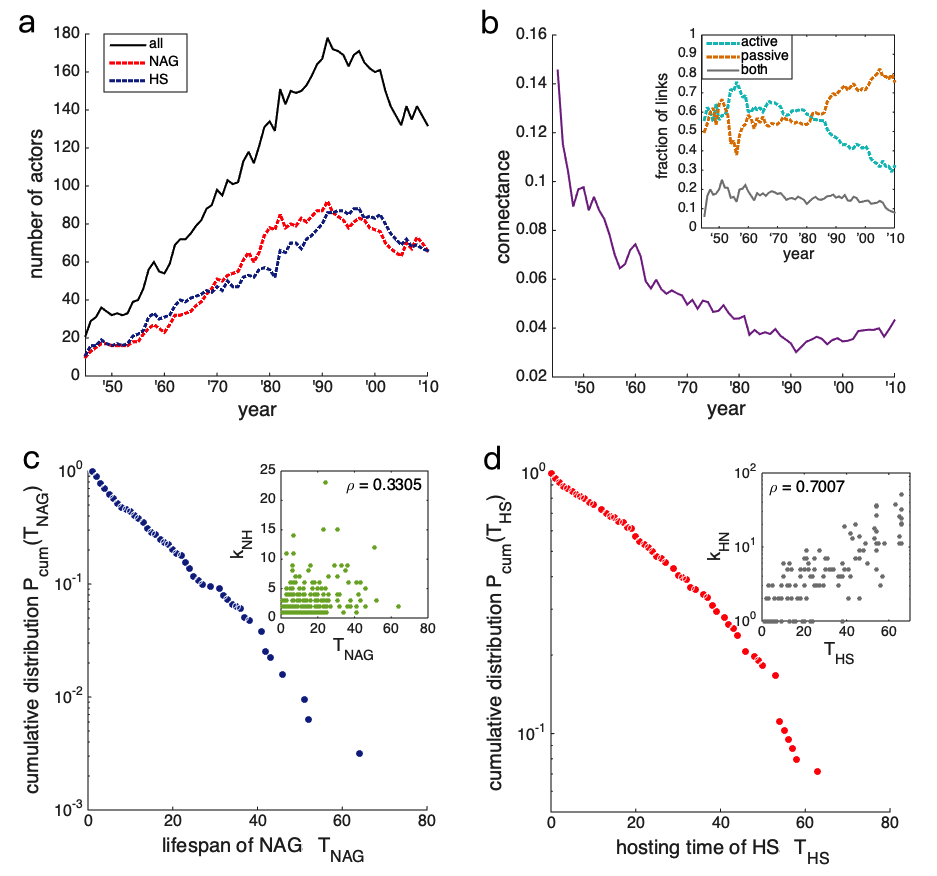}
\caption{\label{NAG_CDF_H_N_INSET} Evolving NAG-HS support network. \textbf{a}, Changing size of the bipartite network over time, in terms of the number of all actors and the components. \textbf{b}, Network connectance, defined as the ratio of the actual number of links against the maximum possible number of links ($n_{NAG}*n_{HS}$). Inset: fractions of links that correspond to active, passive and both types of support, respectively. \textbf{c, d}, Cumulative probability distributions of life spans of NAGs and hosting time of host states (exponential-like distribution). Insets: Correlation of life span or hosting time and the number of attached actors (degree). }
\end{figure}

We focus on the relations between NAGs and external state supporters, extracted from the Dangerous Companions Database covering the time period from 1945 to 2010 \cite{San-Akca:2016}. This considers only violent NAGs (that is NAGs whose operations resulted in 25 or more deaths per year of operation). A HS is a country that provides any type of material support to a given NAG, including safe haven to NAG members, training, or military or financial aid. A temporal network is constructed from these bipartite relations, including both active and passive support. During the time period of study, the network experiences substantial growth with an almost monotonic increase peaking in the early 1990s, corresponding to first years in the post-Cold War era, followed by a decrease in both numbers of nodes and links (Fig. \ref{NAG_CDF_H_N_INSET}a and Fig. S1 in SI), whereas the connectance (link density) shows an opposite trend (Fig. \ref{NAG_CDF_H_N_INSET}b).

First we track the assembly and disassembly of the NAG-HS network, focusing on the relative probabilities of acquisition and disassociation of actors (NAGs and HSs) and links (NAG-HS relations). Then, we show that the network evolves to have 
a particular architecture exhibiting both nested and modular patterns. 
Finally, based on this characteristic architecture, we determine the roles of individual NAGs and HSs.\\


\noindent\textbf{Fitness of actors.} 
We consider two indicators of a NAG's or HS's fitness. The first indicator is based on duration over time, specifically the lifespan of a NAG or the hosting time of a HS (how long the state participates in the supporting activities in any capacity). Generally, we find that these follow an exponential-like distribution, as shown in Fig. \ref{NAG_CDF_H_N_INSET}c and \ref{NAG_CDF_H_N_INSET}d. The second indicator is based on a snapshot of time, specifically the extent of connectivity of the NAG or HS measured by its degree, $k_{NH}$ or $k_{HN}$ respectively. The insets of Fig. \ref{NAG_CDF_H_N_INSET} show the Pearson coefficients $\rho$ between the lifespan/hosting time and node degree ($p<0.0001$ using a two-sided t-test); a NAG tends to endure longer with more supporting HSs, while the hosting time of a HS is more likely to increase the more NAGs rely on it.

These results suggest that ``successful" NAGs---with longer survival time---depend on their ability to extract external support. Likewise, ``active" HSs---states with longer record of support for NAGs---tend to be generalists in that they support more NAGs, compared to the activity duration of ``specialist" HSs.\\


\begin{figure}[t]
\hskip 1cm
\includegraphics[width=0.84\textwidth]{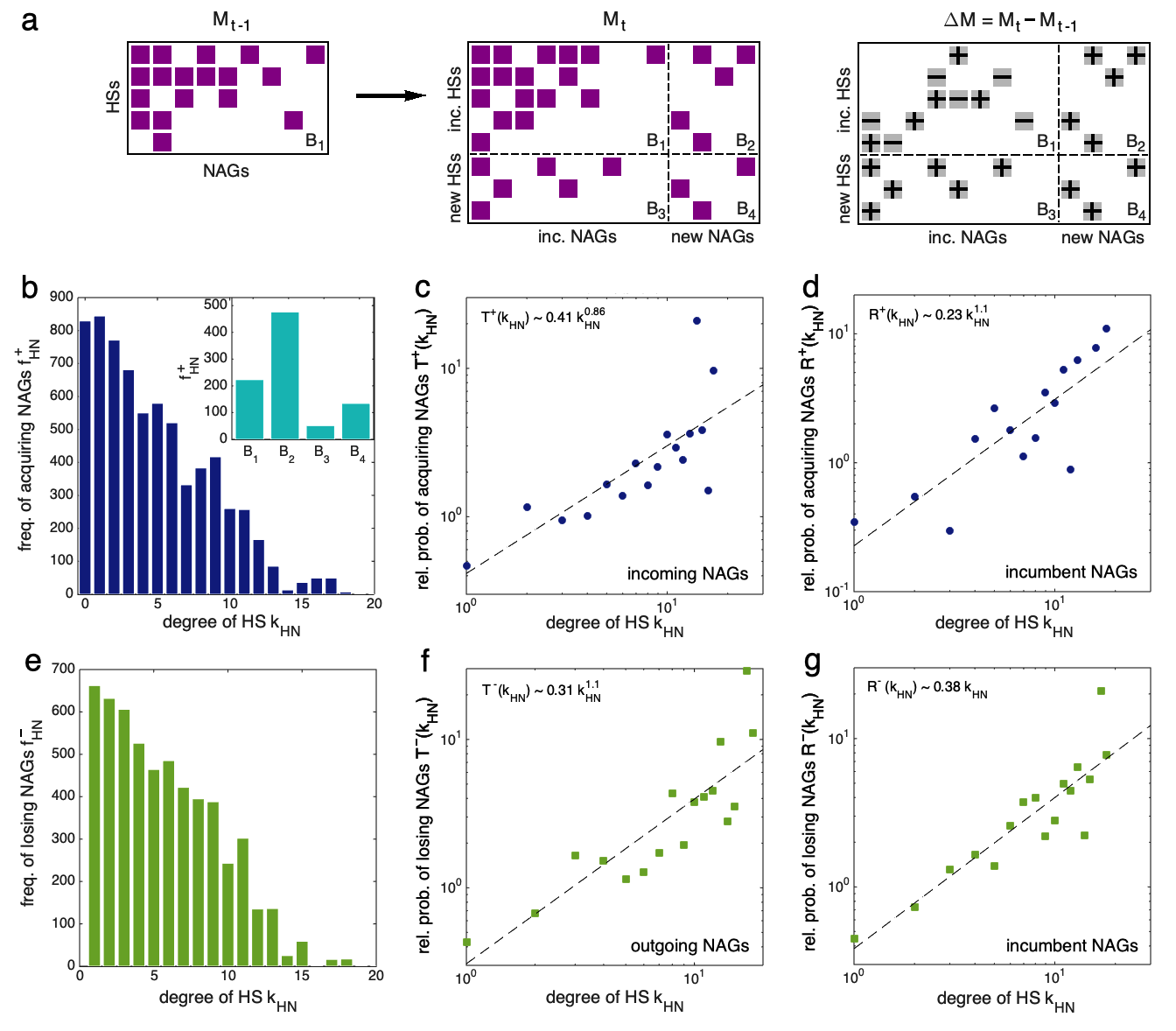}
\caption{\label{NAG_REL_PROB_TK_RK_ALL2}Network assembly and disassembly. \textbf{a}, Link addition (``$+$") and removal (``$-$") as the adjacency matrix evolves from $M_{t-1}$ to $M_t$ in consecutive years. $M_t$ is divided into four blocks ($B_1$ to $B_4$) by separating incumbent and newly joining actors. \textbf{b, e}, Histograms for the frequencies of acquiring and losing links from NAGs by HSs with degree $k_{HN}$ (zero-degree for new HSs). Inset of panel b: frequencies of adding links in the four blocks. \textbf{c}, Relative probability $T^+(k_{HN})$ that an incumbent HS with $k_{HN}$ previous connections acquires links from newly joining NAGs. \textbf{d}, Relative probability $R^+(k_{HN})$ that an incumbent HS acquires new links from an incumbent NAG. \textbf{f}, Relative probability $T^-(k_{HN})$ that an incumbent HS loses links from NAGs that depart the network. \textbf{g}, Relative probability $R^-(k_{HN})$ that an incumbent HS loses links from an incumbent NAG.} 
\end{figure}

\noindent\textbf{Network assembly and disassembly.} To understand how the actors (nodes) and the associated relations (links) attach to and detach from the network, we examine the assembling and disassembling processes that unfold on it. Specifically, we observe how actors acquire (``$+$") and remove (``$-$") counterpart actors by comparing the adjacency matrices in two consecutive years, as illustrated in Fig. \ref{NAG_REL_PROB_TK_RK_ALL2}a. From the perspective of HSs, the histograms show that both attachment and detachment of NAGs occur more frequently at HSs with a lower degree (Fig. \ref{NAG_REL_PROB_TK_RK_ALL2}b and d), but this is biased since the degree distribution of the incumbent HSs is already skewed before those events. It is thus particularly revealing to calculate the relative probability of a HS gaining or losing a link from a NAG, which is equal to the absolute probability divided by the probability from random selection \cite{Newman:2001,Saavedra:2008} (with that any unevenness in the degree distribution is 
offset), over the entire time span of 66 years. A nonrandom distribution of relative probability is observable if there is an intrinsic bias in the attaching or detaching process.  

Indeed, both attachment and detachment of NAGs are found to occur preferentially at incumbent HSs of higher fitness as measured by their degrees. First, we consider the attachment of NAGs, that is, the relative probability for an incumbent HS to acquire links from newly arriving NAGs (NAGs that start to receive support from at least one HS) as a function of the degree of the HS, denoted $T^+(k_{HN})$, and find that a newcoming NAG is more likely to attach to a generalist (high-degree) HS. Explicitly, the relative probability $T^+(k_{HN}) = P^+(k_{HN})/P_0^+(k_{HN})$ is defined as the ratio of the absolute probability $P^+(k_{HN})$ that a NAG joining the network in year $t$ connects to an incumbent HS that has already been supporting $k_{HN}$ NAGs divided by the probability $P_0^+(k_{HN})$ that a HS with degree $k_{HN}$ is uniformly selected at random (see Methods). $T^+(k_{HN})$ is shown 
in Fig. \ref{NAG_REL_PROB_TK_RK_ALL2}c, which is approximately fit by $T^+(k_{HN})\sim k^{0.86}_{HN}$, indicating a 
(slightly sublinear) preferential attachment (PA) process \cite{Yule:1925,Barabasi:1999}.
Turning to detachment, we find the same tendency holds. Specifically,  
a NAG departing the supporting network (a NAG has no support from any HS or has been dismissed) is more likely to detach from a counterpart HS supporting more NAGs, as shown by the linearly increasing relative probability $T^-(k_{HN})\sim k^{1.1}_{HN}$ with the degree of the latter (see Fig. \ref{NAG_REL_PROB_TK_RK_ALL2}f). This (slightly superlinear) preferential detachment (PD) is somewhat counter-intuitive as it suggests that the network tends to disassemble at higher-fitness nodes. 

Second, 
besides the contribution from newcoming NAGs, an incumbent NAG (a NAG that receives support from at least one HS) of the supporting network may also add or delete links to another incumbent HS over time. To include such cases, 
we measure the relative probability $R^+(k_{HN})$ (slight superlinear) that an incumbent HS that has $k_{HN}$ links in the year prior to $t$ will acquire links from incumbent NAGs in year $t$, excluding those from newcoming NAGs (Fig. \ref{NAG_REL_PROB_TK_RK_ALL2}d). Similarly, Fig. \ref{NAG_REL_PROB_TK_RK_ALL2}g shows the case for incumbent HSs losing links from incumbent NAGs (remaining in the network), as indicated by $R^-(k_{HN})$ (linear). All these attachment and detachment show that NAGs interact preferentially with HSs of greater fitness. We can also observe the tendencies from the perspective of NAGs, in terms of acquiring and losing links from HSs. Yet, the dependencies are more obscure (though the overall positive dependencies are still observable), due to the paucity of data points at some degrees (see Fig. S5 in SI). 

The above analyses thus show that the assembly and disassembly of the supporting network, including joining and departure of NAGs, and rewiring between incumbent actors, proceed with propensities that are essentially dependent on nodal degree, which seems to reflect the fitness of the HS supporters. 
However, these analyses also show two conflicting processes: 
while the attachment is driven in attempt to maximize individual fitness, having higher fitness can also lead to higher possibility of losing counterpart actors, which has not been observed in other bipartite network disassembly \cite{Saavedra:2008}. 
As discussed later, this could reflect the fact that an actor with many alliances may benefit from having fewer interactions of higher quality. 
\\

\begin{figure}[tb]
\hskip 0.3cm
\includegraphics[width=0.95\textwidth]{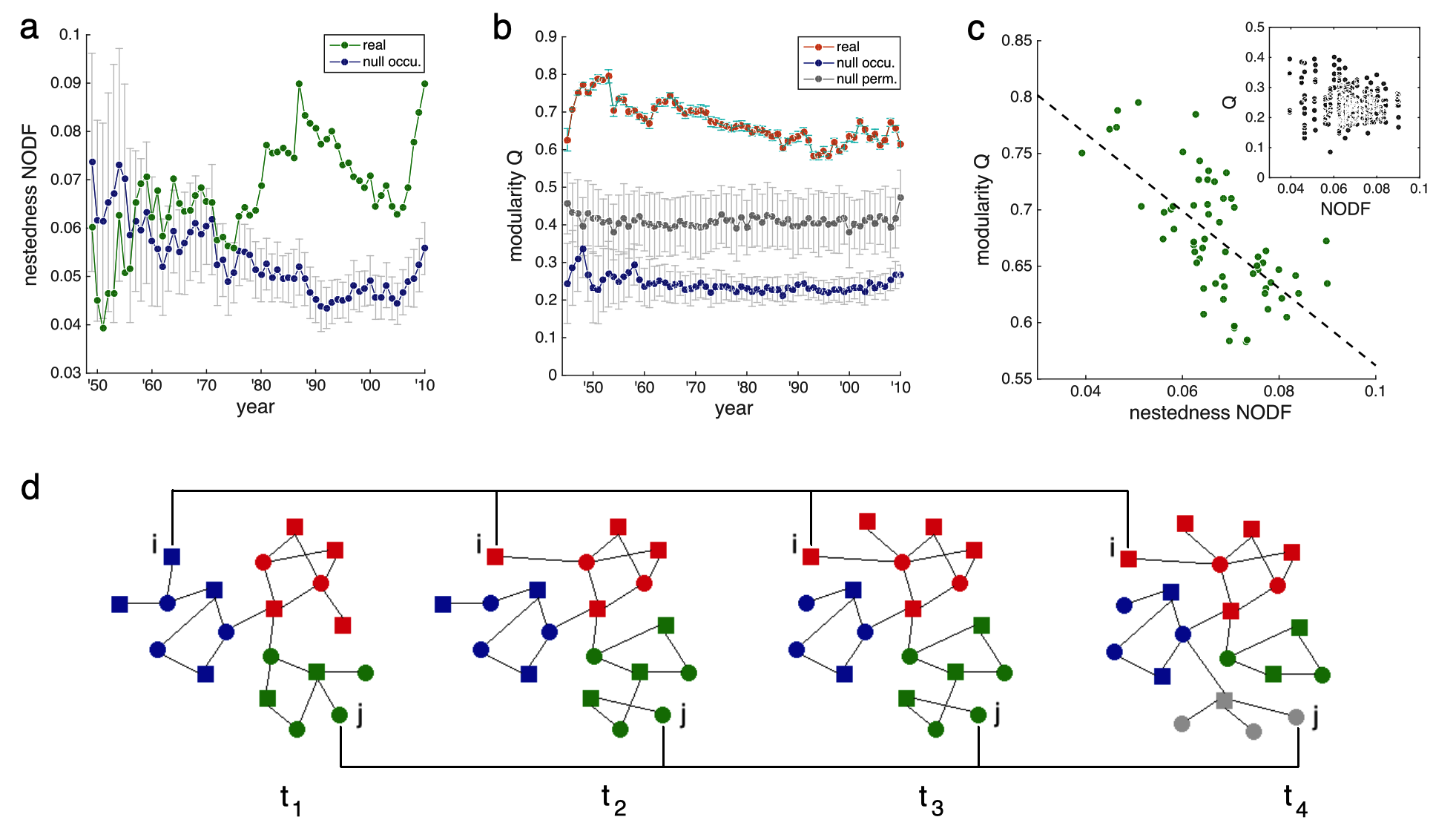}
\caption{\label{NG_NODF_MOD_LL_CORR}Evolution of structural measures of the temporal NAG-HS network: (a) nestedness NODF and (b) modularity Q. The nestedness is measured for the 5-year aggregated networks in a sliding window; the modularity is obtained by applying a temporal detection algorithm (note that the optimization algorithm can result in slightly different partitions of the same network). The results for the empirical network are compared with corresponding null models: I. average occupation probability (labeled ``null occu.") and II. random slice permutation (labeled ``null perm."). Confidence interval of one standard deviation is used. (c) Negative correlation of modularity and nestedness in the NAG-HS supporting network, in comparison to the random relation between the two measures for the null (occu.) model in the inset. (d) Temporal 
community detection algorithm. The example temporal bipartite network (circles for NAGs and squares for HSs) consists of four time slices that are linked by connecting nodes in one slice to themselves in the adjacent slices (demonstrated for nodes $i$ and $j$). The partitioning is based on the entire temporal network and the yearly modularity is then obtained according to the partition in each slice (each color representing one module). }
\end{figure}

\noindent\textbf{Nested structures.} Despite significant variations in the network size (numbers of actors and links), the network exhibits a robust nested and modular architecture over a long time span. The measure of nestedness simply reflects to what extent actors share the same set of partners. It is calculated with the NODF metric (``nestedness metric based on overlap and decreasing fill") \cite{AlmeidNeto:2008}. To show how it evolves, we calculate the nestedness of the aggregated network in a 5-year sliding window, as shown in Fig. \ref{NG_NODF_MOD_LL_CORR}a (this is to avoid statistical fluctuation due to the sparsity of entries in single yearly network slices). Between 1945 (end of World War II) and 1975 (end of Vietnam War), the nestedness keeps at a level that is comparable to that of a random network. Starting from 1975, the empirical nestedness score exceeds significantly the random scores, peaking in 1987 (towards the end of the Cold War). The increase parallels the cumulative broadening of degree distribution due to the preferential attachment and detachment processes (see the comparisons in Fig. S4 in SI). This is then followed by a dip and then upswing back to a similarly high level in 2010. For comparison, random networks are used as the null model, where the probability of each entry of the adjacency matrix being occupied is the average of the occupation probabilities of the row and column \cite{Bascompte:2003}. The significantly high nestedness implies that actors (say NAGs) are more likely to share common partners (say HSs) which tends to make these common partners generalists. This renders a highly unbalanced structure of relationships: specialists (say NAGs only focusing on few HSs) are more likely attached to generalists (HSs supporting many NAGs) rather than other specialists (HSs supporting only few NAGs). 


A heuristic explanation can be suggested for the formation of the nested architecture, following the principle of fitness maximization in ecology \cite{Suweis:2013,San-Akca:2009}: For those NAGs and HSs maintaining mutualistic relations, the successful military action taken by a NAG may improve the status of the counterpart HS and thus increase the state's propensity and willingness to support more NAGs. 
And the successful NAG may also attract support from more HSs intending to confront their rivals in this way. Similarly, for parasitic NAG-HS relations, the feasibility of one NAG gaining resources from a HS may encourage more NAGs to attach to that state.  
This mechanism is weakened by the detaching events, which however occur much less than the attaching events up to the middle of 1990s (see the number of links in Fig. S1a). This has led to ``net" preferential attachment.  

A special note should be made to the indirect benefit for the NAGs in the nested configuration \cite{Bastolla:2009,Maoz:2012,Phillips:2018}. The NAGs that engage in mutualistic support from the same HSs may avoid or reduce possible competition between these NAGs for territory or resources. Since successful military actions by a given NAG may improve the willingness of HS to support additional NAGs, NAGs can be thought to indirectly contribute to each other's support levels. This complies with the general ecological principle of mutualism \cite{Bastolla:2009}: a more nested architecture is able to pack more actors in the community by minimizing competition among like-type actors. This higher capacity marks a more structurally stable architecture \cite{Rohr:2014}. \\

\begin{figure}[t]
\centering
\includegraphics[width=0.98\textwidth]{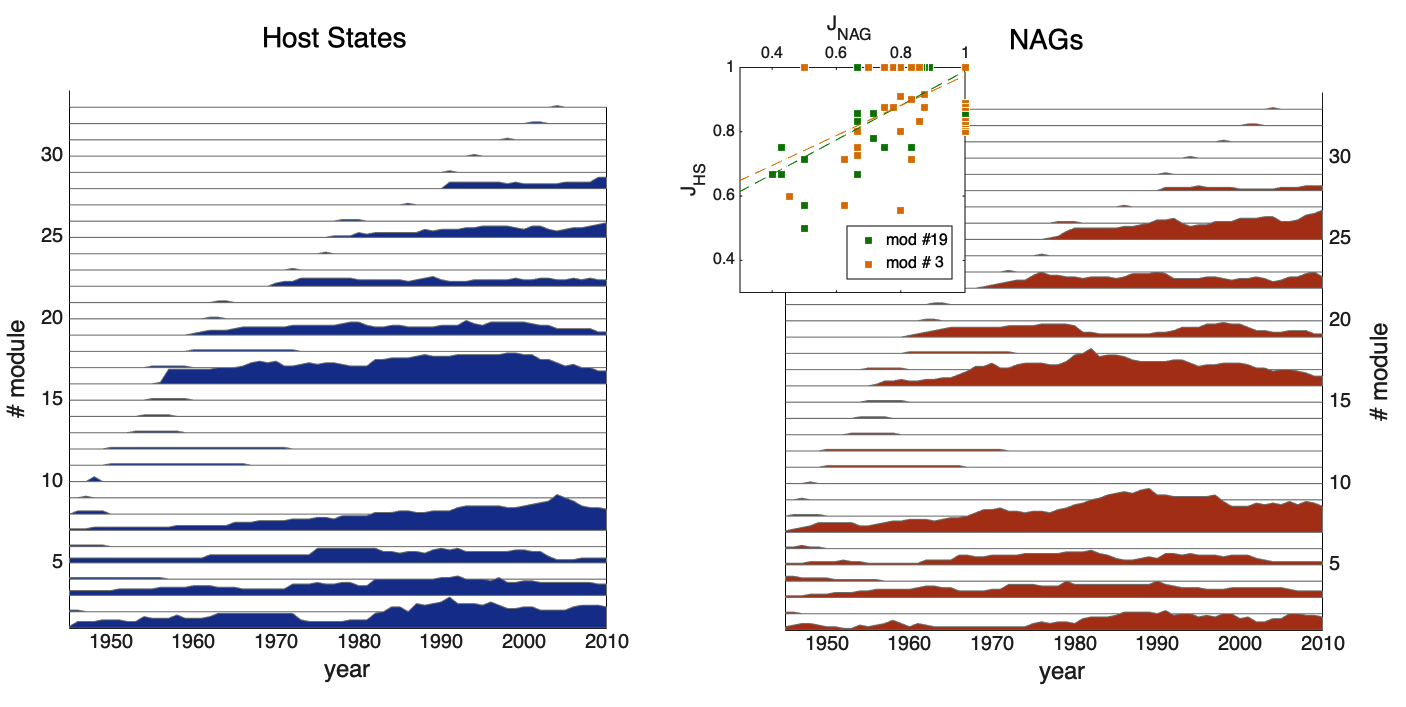}
\caption{\label{NG_MOD_YR_HN}Temporal modular structure in the (a) HS and (b) NAG guilds. The height of the curve represents the number of actors in the submodule. The modules are numbered (vertical axis), among which 9 modules contain more than 3 different actors in each while the other 24 are small groups or individual actors. Inset: Positive correlation of the Jaccard indices of NAG- and HS-submodules of two of the major modules shows coherent fluctuation in their membership (Pearson coefficient $\rho = 0.71$ with $p = 6.1 \times 10^{-9}$ (two-sided t-test) for mod \#19; $\rho = 0.56$ with $p = 9.9 \times 10^{-7}$ for mod \#3).}
\end{figure}


\noindent\textbf{Modular structures.} Concurrently, the network exhibits a nontrivial modular structure, in which actors are more densely connected to partners within the same module than across modules. The modularity measure is usually calculated by seeking a partition of an individual network that maximizes a modularity quality function \cite{Newman:2006}. However, for this temporal network, partitioning the yearly slices of the temporal network individually would lose the inter-slice relation, making the modules in successive years independent of each other. Here we use a temporal community detection algorithm based on Markov stability, proposed by Mucha, et al \cite{Mucha:2010}. Instead of partitioning separate time slices, this algorithm relies on the systematic optimization of a single quality function for all network slices in the 66 years. It has the merit of allowing the identification of the same module through all years and thus tracking the evolution of modules \cite{Fortunato:2010}.  

The modularity of the network each year as measured via the metric $Q(t)$ is significantly higher compared to that of corresponding null models, as shown in Fig. \ref{NG_NODF_MOD_LL_CORR}b (note that the partition is not unique). There are two appropriate null models: 1. randomly permuted yearly slices (see Fig. S11a) and 2. randomly shuffled within-slice links that maintain the same degree sequence (see Fig. S11b). The comparisons demonstrate that both intra-slice linkage and the temporal order of network slices contribute to the nontrivial modular structure \cite{Bassett:2011}. 
The modularity value reaches a minimal value around 1990, a time point marking the end of the Cold War.  


\begin{figure}[t]
\hskip -0.2cm
\includegraphics[width=1.02\textwidth]{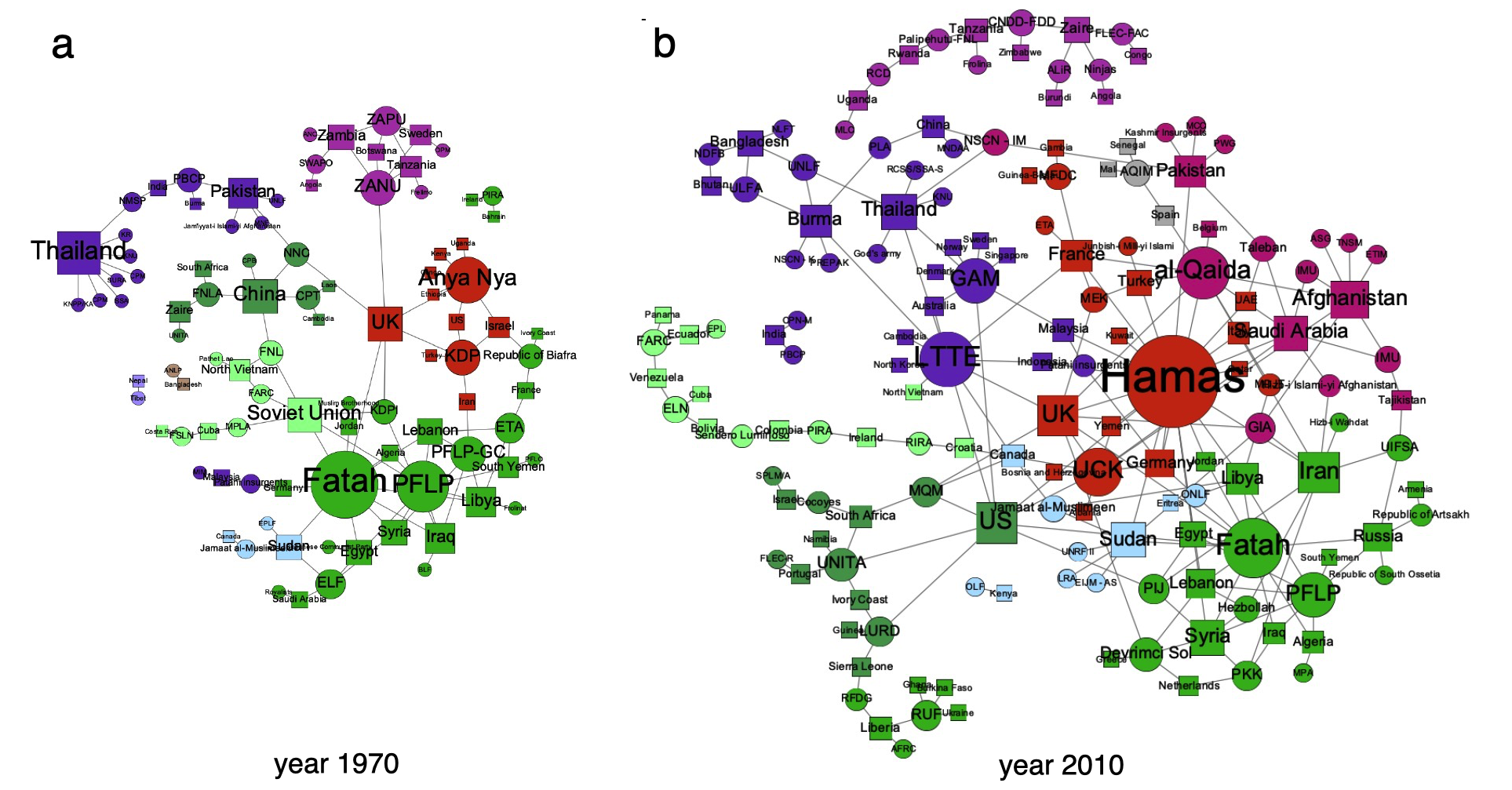}
\caption{\label{Y1970-2000F} Partitioned NAG-HS networks for representative years (a) 1970 and (b) 2000. Nodes are coloured differently according to their bipartite module-membership, but the colour of the same module does not alter for different years. NAGs are represented by circles and HSs by squares.
}
\end{figure}


We then seek a representative partition of the network. This is realized by using a stabilizing technique for robust module detection \cite{Bassett:2013} and maximizing the Adjusted Mutual Information (AMI) (see Methods and Sec. 2 in SI). Our more detailed analyses are then based on this unique representative partition. The supporting network over the 66 years is partitioned into 33 modules, in which 9 major modules contain more than 3 different actors in each and the other 24 modules are separate nodes or tiny groups. In Fig. \ref{NG_MOD_YR_HN}, we show the sizes (number of actors) of modules in all years in the HS- and NAG-guilds (the NAG- and HS-submodules of each bipartite module), respectively (see also the modular structures in representative years in Fig. S9 in SI). The contrast is sharp: while the major modules can persist for significantly long time spans, the others are merely temporary. 

While the 24 temporary modules are almost constant in their membership, all the major 9 modules have experienced significant turnovers (see module membership in 1970 and 2000 shown in Tab. S1 and S2 in SI). This is consistent with the general tendency found in the unipartite coauthorship and the phone-call networks \cite{Palla:2007}: small modules survive when they are stationary in the membership, while large modules survive if they experience sufficient changes. For the 9 major modules, we further observe that the fluctuations in the NAG- and HS-submodules are coherently correlated. Quantitatively, we calculate the temporal Jaccard index to show the stationarity of the membership of the submodule, which compares the contents of in the successive submodules 
\begin{equation}
\label{eq:jac}
J_{\alpha}(t,t-1)=\frac {|A_{\alpha}(t)\cap A_{\alpha}(t-1)|}{|A_{\alpha}(t)\cup A_{\alpha}(t-1)|}
\end{equation}
where $A_{\alpha}$ denotes the members of a specific submodule in focus \cite{Palla:2007,Bassett:2013}. Thus, the Jaccard index describes the similarity of successive modules (auto-correlation), ranging from 0 (no common member) to 1 (identical). For all submodules of the 9 major modules, the Jaccard indices averaged over the active years turn out to be high (see values in Fig. S6 in the SI), indicating high cross-time stability of these modules. In individual modules, for 7 of the 9 major modules, the temporal Jaccard indices of the corresponding NAG- and HS-submodules, $J_{NAG}(t)$ and $J_{HS}(t)$, are positively correlated ($p<0.1$; two-sided t-test), as illustrated for the two modules with the highest correlation coefficients in the inset of Fig. \ref{NG_MOD_YR_HN} (see Fig. S7 in the SI for all major modules). This implies that the constituents of most major bipartite modules are co-changing: NAGs and their attached states tend to join or leave a module together.

Viewing the nestedness and modularity measures jointly, we find a clear negative correlation between them through the 66 years, as shown in Fig. \ref{NG_NODF_MOD_LL_CORR}c. This is strikingly different from that observed for the null model. Heuristically, on one hand, NAGs may favor a nested structure by attaching to common HSs so that the HS may increase the willingness to provide more support. On the other hand, the benefits of support could be diluted if the difference between the traits (including geographic, political and religious factors) of a pair of NAG and HS is too significant, which may lead them to fall into different modules. This is consistent with the ecological principle that at any given time, the network tends to find an optimal structure balancing nestedness and modularity, which trades off benefits against disadvantages \cite{Bastolla:2009,Suweis:2013,Cai:2020}. \\

\begin{figure}[t]
\hskip 0.2cm
\includegraphics[width=0.95\textwidth]{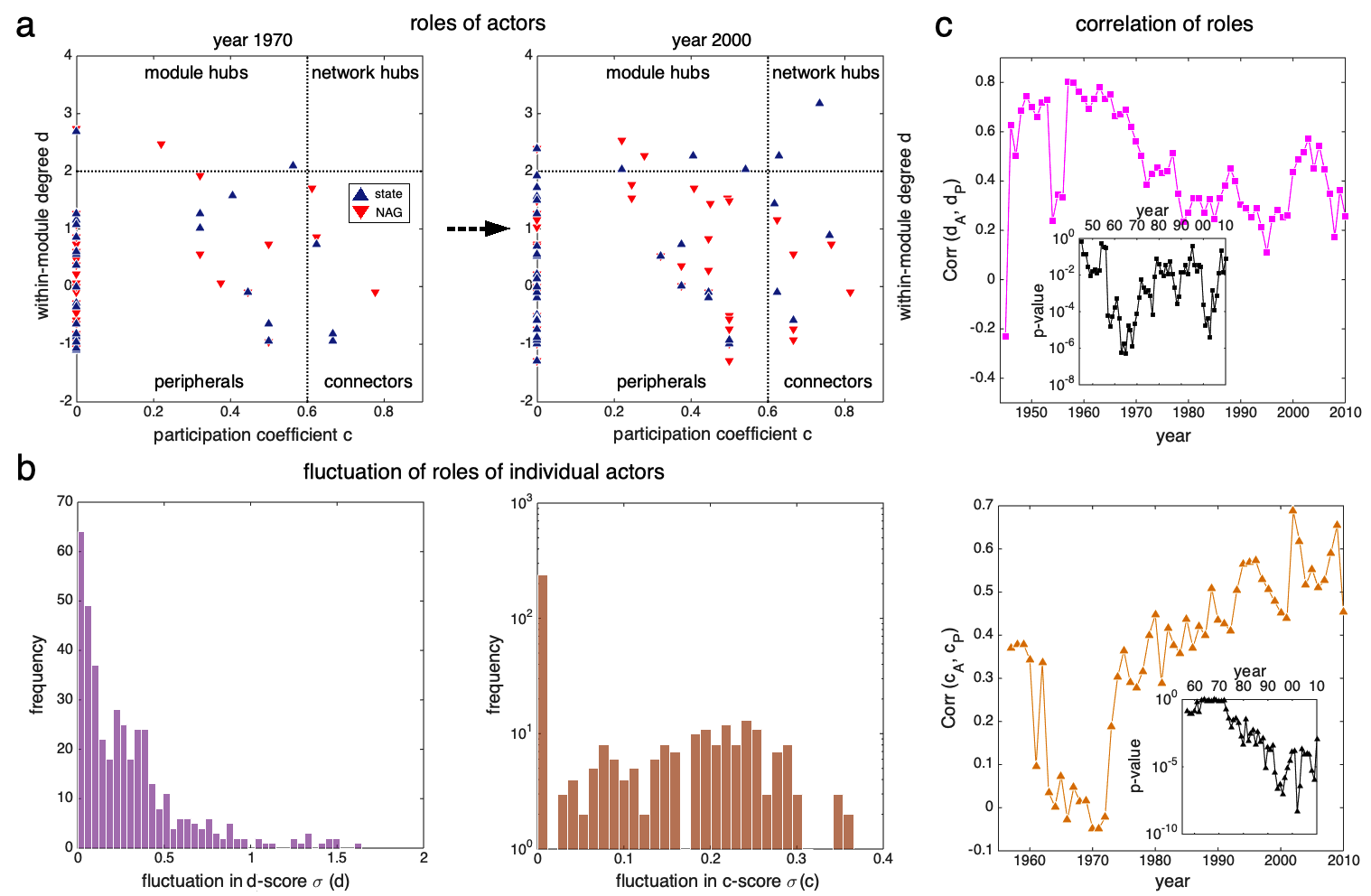}
\caption{\label{NAG_ROLES_ALL2} Roles of actors. \textbf{a}, Standardized within-module degree $d$ and participation coefficient $c$ for the network in year 1970 and 2000, respectively. According to the two scores, any node falls into one of the four predefined categories. \textbf{b}, Histograms of fluctuation in d- (left panel) and c-scores (right panel) measured by the standard deviation. \textbf{c}, Correlation of temporal d-scores (upper panel) and c-scores (lower panel) in the active and passive support subnetworks, for the actors that are involved in both relations. Insets: p-values of the temporal correlations. }
\end{figure}

\noindent\textbf{Roles of actors.} Based on the representative partition, we may further identify the roles of actors in the network setting. We calculate two characteristics of each actor, the standardized within-module degree $\{d_i\}$ and the participation coefficient $\{c_i\}$ \cite{Guimera:2005}. The c-score is 0 if all the node's links are within its own module and approaches 1 if they are distributed evenly among modules. The two scores characterize the contribution of a node to the degree distribution (thus to the nestedness) and modular structure. Then the roles can be classified into four categories: peripherals, module hubs, connectors and network hubs. Figure \ref{NAG_ROLES_ALL2}a shows the two scores of all actors in the networks in two typical years 1970 (I) and 2000 (II), respectively. It is notable that although the number of actors, including both NAGs and HSs, has increased from 98 to 160, the distribution of roles of all actors has remained very similar (mean value: $\langle d^I \rangle = \langle d^{II} \rangle = 0$, $\langle c^I \rangle = 0.1086$ vs $\langle c^{II} \rangle = 0.1402$, standard deviation: $\sigma(d^I) = 0.9250$ vs $\sigma(d^{II}) = 0.9582$, $\sigma(c^I) = 0.2133$ vs $\sigma(c^{II}) = 0.2348$). For each actor, we measure the fluctuation of its yearly role scores by the standard deviation throughout its lifespan (for NAG) or hosting time (for HS). As shown in Fig. \ref{NAG_ROLES_ALL2}b, most actors turn out to be consistent in their d-scores, with $82\%$ of all actors  experiencing fluctuations less than 0.5. For the c-score, $67\%$ of the actors remain the same or have a fluctuation less than 0.1. Most actors have thus kept their within-module degree (reflecting its fitness) unchanged and meanwhile stayed similar in the role of connecting different modules. In this sense, the roles of the actors (being hubs or peripherals) tend to be maintained in the network evolution regardless of addition and deletion of actors. 

Analogous to ecological mutualistic and parasitic relations, the NAG-HS network consists of active and passive support relations, which comprise two subnetworks, each containing only one type of support. Now we analyze the roles of the actors that are involved in both relations. We project the partition of the whole network onto the subnetworks (so the actor's module-membership in the subnetwork is determined by the partition of the whole network).
We calculate the d- and c-scores of actors in each year in each of the subnetworks, respectively, and then the Pearson coefficients between the yearly d-/c-scores of all actors for the two subnetworks, as shown in Fig. \ref{NAG_ROLES_ALL2}c. Except a few years, the correlation of the d-scores is significantly high in the majority of the 66 years $(p<0.01$; two-sided t-test). Similarly, the correlation of c-scores shows an increasing tendency over time from 1975 on and becomes significantly high from 1992 on $(p<0.01)$. 

The correlation indicates that an actor involved in both active and passive relations generally plays a consistent role in both subnetworks. Concretely, if the actor plays a role as a peripheral/connector/module-hub/network-hub in one subnetwork, it tends to play a similar role in the other subnetwork. In terms of ecology, this means that if a NAG relies heavily on mutually beneficial relations with supporting states, it is very likely to develop substantial parasitic relations as well, such as extracting funds, transporting weapons or finding safe heavens without the approval or notice of the HS \cite{San-Akca:2009,Phillips:2015}. It is notable here that a HS supporting a set of NAGs actively may support another set of NAGs passively. For example, the US provided active support to the Mujahedin in Afghanistan in the 1980s, and it was exploited otherwise by the IRA, PLO, and Hamas during the same period.\cite{guelke1996, levitt2008, abu1993}

\section*{Discussion}

As for any ecological system, identifying the resources that cultivate the actors is crucial for understanding the actors' life cycles and their collective behaviours. Although NAGs are not persistently owned or controlled by any state, they acquire resources from selected states in a nonrandom manner. The support network is self-organized through a process where every actor aims to maximize their individual fitness \cite{Suweis:2013,Cai:2020}, which is known as ``cumulative advantages" \cite{deSollaPrice1976,Perc2014}. This is reflected by the preferential selection of partnerships and results in characteristic network patterns. This means that NAGs select HSs that are most likely to provide support for longer periods of time. For a similar reason, 
long-lasting NAGs tend to attract HSs, which tend to cluster around the same NAGs.  


The tendency of attaching to higher degree nodes is consistent with other network formation processes, such as in the unipartite scientific collaborations, the bipartite plant-pollinator relations \cite{Bascompte:2003}, or the designer-contractor partnerships \cite{Saavedra:2008}. Yet the generalized preferential detachment in the disassembly of the NAG-HS network is unique. It means that an actor with more associated partners (if mutualistic) or exploiters (if parasitic) may be more likely to break a link, which is most probably sub-optimal. 
This may be due to the fact that an actor with more links tends to be more selective over time to minimize its dependence on others. Since supporting a NAG is an act of animosity and might attract retaliation, limiting interactions is the smart strategy to pursue by HSs. By the same token, if we consider that HSs might also constrain the operations of NAGs, the latter would try to keep their links limited. For example, Arafat was particularly careful about getting support from Syria and Egypt in the 1970s since both of these countries had strong leaders who prevented Palestinian infiltration to Israel from their own territory, for fear of Israeli retaliation. Instead he opted to form strong ties with Jordan (in the late 1960s) and with Lebanon in the 1970s and early 1980, because he believed that these states had weak governments that could be exploited \cite{Sayigh:1997,Rubin:2003, maoz2006}. 


Despite continuous attachment and detachment, the self-organized nontrivial architecture persists over a wide time span. Moreover, the independent measures of nestedness and modularity turn out to be negatively correlated in the NAG-HS supporting network, but no correlation appears in the null model. While nestedness can result from the compound effect of PA and PD, modularity seems to be the consequence of matching of actors' traits, which are determined by geographical, political and religious factors. A similar negative correlation is also observed between the dyadic measures of a large ensemble of ecological mutualistic networks, across a wide range of geographical factors and constituents \cite{Cai:2020}. The emergence of a nested, modular structure and their negative correlation suggests that this network of military support and mutualistic ecological networks may share a common ecological principle, that is, optimizing fitness in light of tradeoffs. 

The temporal community detection algorithm confirms the existence of an almost unique partition of the support network (see Sec. 2 in SI). Most importantly, this approach is able to track the temporal changes of modules. Unlike the modules in the US congress roll call network \cite{Mucha:2010}, partitioned by using the same algorithm, where a range of duration periods of modules appear, the modules in the NAG-HS supporting network are either long-lasting ones containing a variety of members or short-lived ones consisting of less than three pairs of interacting actors. This suggests that a module has to contain sufficiently many actors and internal interactions to persist; it cannot survive on only few isolated interactions.

Geographic similarity is shared by most actors in each of the nine established modules (see Tab. S1 and S2 in SI for contents in two representative years), as most prominently 
demonstrated by the HSs 
and would thus give credits to regional studies that focus essentially on internal dense interactions \cite{San-Akca:2016,San-Akca:2009,Phillips:2015}. However, it is notable that remote actors beyond the region are also often present in the same module. Most of them are major powers with global force and financial projection capacity, connecting several modules, including US, UK, Russia (Soviet Union) and China. These were related to the Africa- or Mideast-centered regions (see Fig. \ref{Y1970-2000F} for the partitions in 1970 and 2000). In addition, a few states may have transregional influences due to specific historical reasons, for instance, the leaders of GAM, a separatist group in Indonesia, are known to have lived in and operated from Sweden for most of the 80s and 90s before its dissolution in 2005 \cite{Schulze:2004}. This suggests that the temporal partition of modules may reveal deeper relations of intimately connected actors beyond the simple methodology of regional division and thus justifies the modules as proper study objects.


Revealing the nontrivial structure enhances our understanding of the roles of the actors. It allows us to capture the manner in which certain NAGs act as connectors, and in which some states form large connecting hubs for multiple NAGs. This may open the door for studying policies to cut off key actors that may cause malfunction of actors only within a module, or lead to cascades that would have extendable influence on multiple modules. These research avenues can benefit from the large body of recent studies on the percolation and cascading failure in modular and/or nested networks \cite{Saavedra:2011,Memmott:2004,Burgos:2007,KaiserBunbury:2010,Gleeson:2008,Snyder2020}, as well as on their structural stability and capacity \cite{Rohr:2014}.   


In all, quantifying the ecosystem of NAG-HS interactions provides a unique and important perspective towards the comprehension of the origin, evolution and termination of NAGs, the states supporting them, and their collective behaviours. Ecological organizing principles, originally discovered in natural systems, seem to be shared by this military network, as well as a variety of previously studied socio-economic networks \cite{May:2008,Saavedra:2009}. We believe further investigations of the NAG-HS network will be fruitful for the domain of warfare and conflicts as well as developing theories to explain the origin of the ubiquitous network patterns beyond the domain.


\section*{Methods}

\textbf{Dataset of NAGs.} Our study objects are the bipartite adjacency matrices between NAGs and their attached states of support, constructed from the Dangerous Companions Database \cite{San-Akca:2016}. The dataset covers the period between 1945 and 2010, which introduces a profile for each NAG with information on the foundation year, objectives and ideational characteristics. All NAGs are engaged in a violent conflict against one or more governments and only conflicts that caused more than 25 battle-related deaths (BRD) are collected, namely the major ones. 
A HS can provide multiple types of support to a NAG
each of which is classified as active if it is intentional or passive (de facto) if it is inadvertent. \\  


\noindent\textbf{Relative probability.} We calculate the relative probability $T^+(k_{HN})$ that a NAG added in year $t$ connects to an existing state which is connected to $k_{HN}$ NAGs in year $t-1$. The relative probability $T^+(k_{HN}) = P^+(k_{HN},t)/P_0^+(k_{HN},t)$ is defined as the ratio of the absolute probability $P^+(k_{HN},t)$ that a NAG connects to a HS with degree $k_{HN}$ against the the probability $P_0^+(k_{HN},t)$ that a HS with degree $k_{HN}$ is uniformly selected at random, where $P_0^+(k_{HN},t) = n_s(k_{HN},t)/N_s(t)$ with $n_s(k_{HN},t)$ being the number of HSs with degree $k_{HN}$ prior to the addition of this NAG and $N_s(t)$ being the number of all HSs. Then $T^+(k_{HN})$ can be estimated by making a histogram of the degrees $k_{HN}$ of the HSs to which each NAG is added, in which each sample is weighted by a factor of $N_s(t)/n_s(k_{HN},t)$ \cite{Newman:2001}. In the random case, $T^+(k_{HN})$ should equal 1 for all $k_{HN}$. If it is a nonconstant function of $k_{HN}$, some bias is involved in the attachment process. In particular, a linearly increasing function corresponds to the standard preferential attachment (PA); while a general monotonically increasing function corresponds to generalized PA. Similarly, we calculate the relative probability $T^-(k_{HN})$ that a NAG detaches from a HS with degree $k_{HN}$. Furthermore, the relative probability of a HS acquires or loses a connection from an incumbent NAG, namely, $R^+(k_{HN})$ and $R^-(k_{HN})$. From the perspective of NAGs, we can also calculate the relative probability that a NAG acquires or loses a connection from a HS (see Fig. S5 in SI). \\

\noindent\textbf{Temporal detection of modules.} We use a temporal community detection algorithm, GenLouvain, based on the Markov stability, proposed by Mucha, et al \cite{Mucha:2010}. The merit of the temporal algorithm is its ability to find a systematic optimal partition of the network by optimizing a single quality function throughout all years. Note that by using this optimization algorithm, there can still be arbitrariness due to different partitions $\{P_s\}$ of the same network that give comparable modularity values close to the optimum. A specific $s$th partition is denoted by $P_s(t)=\cup_i M_{s,i}(t)$, where the subscript $i$ denotes the $i$th module and $t$th year. 


We then seek a representative partition if the network has a truly nontrivial modular structure (for details see Sec. S2 in SI). This is realized by using a further stabilizing technique for robust module detection \cite{Bassett:2013}. We construct a nodal association matrix $T$, each entry of which is the frequency of a pair of actors being partitioned in the same module according to the partitions $\{P_s\}$ (only statistically significant contributions are retained). We then partition this matrix $T$, which reveals the network modular structure from all partitions $\{P_s\}$. It turns out that all partitions $\{\tilde{P}_s\}$ of $T$, after applying this technique, are almost identical to each other, indicating the existence of a highly robust modular structure of the studied network. Finally, to remove the remaining slight uncertainty, we select a representative partition $\tilde{P}^*$, which has a maximum average Adjusted Mutual Information (AMI) with all other partitions $\{\tilde{P}_s\}$ \cite{Vinh:2010}.\\ 

\noindent\textbf{Roles of actors.} We calculate two characteristics of each actor, the standardized within-module degree and the participation coefficient \cite{Guimera:2005}
\begin{eqnarray}
\label{eq:zscs}
d_i&=&(k_{is}-\langle \{k_{is}\} \rangle)/\sigma(\{k_{is}\}) \\
c_i&=&1-\sum_{t=1}^{N_m}k_{it}/k_i
\end{eqnarray}
where $k_i$ is the degree of nodes $i$, $k_{is}$ is the within-module degree (number of links of node $i$ to other nodes in its own module $s$), $\langle \{k_{is}\} \rangle$ and $\sigma(\{k_{is}\})$ are the average and standard deviation of within-module degree of all nodes in $s$, and $k_{it}$ is the number of links from node $i$ to nodes in module $t$.

\section*{Data Availability}
The ``Dangerous Companions" Dataset used in this study is available at http://armedgroups.net.

\begin{sloppypar}

\section*{Code Availability}
Codes for analyzing the network are available upon request. The code for measuring nestedness from the library ``BiMat" is available at https://bimat.github.io. The code for temporal community detection from the library ``GenLouvain" is available at https://github.com/GenLouvain/GenLouvain. 

\end{sloppypar}




\section*{Acknowledgements}
This work is supported by U.S. Army Research Office MURI award No. W911NF-13-1-0340 and Cooperative agreement No. W911NF-09-2-0053, by NSF grant DMS-1817124, by DARPA award W911NF-17-1-0077, and by Minerva grant W911NF-15-1-0502. The authors are solely responsible for the views and analyses in this study.

\section*{Author contributions}
W.C., Z.M. and R.D. conceived the study, B.S. and Z.M. provided the data, W.C., J.S. and G.G. conducted the statistical analyses, all authors analyzed the results and wrote the manuscript.

\clearpage
\newpage
\pagebreak

\setcounter{equation}{0}
\setcounter{figure}{0}
\setcounter{table}{0}
\setcounter{page}{1}
\makeatletter
\renewcommand{\theequation}{S\arabic{equation}}
\renewcommand{\thefigure}{S\arabic{figure}}

\noindent\textbf{\LARGE Supplementary Information: Quantifying the Global Support Network for Non-State Armed Groups (NAGs)}\\

\renewcommand{\thetable}{S\arabic{table}}
\renewcommand{\theequation}{S\arabic{equation}}

We present here supplementary information on the NAG dataset, the basic statistics of the bipartite NAG-HS supporting relations and the detailed algorithm for robust partitioning of the temporal network.

\section*{S1. Dataset and More Structural Properties}
The Dangerous Companions Database covers major violent conflicts that have caused more than 25 battle-related deaths (BRD) in the period between 1945 and 2010 \cite{San-Akca:2016S}. Each case in the data set is a triad relation that involves a nonstate armed group (NAG), a supporting host state (HS) and a targeted state (a state suffered violence from one or more NAGs). The dataset is a substantial extension of supporting relation section in the UCDP/PRIO Armed Conflict Database (Gleditsch et al. 2002, Themnér  Wallensteen 2011). We focus on the dyadic supporting relations between NAGs and HSs, from which the supporting network is established. Each of such relations can be consisted of multiple types, each being either active or passive, depending on whether it is intentional on the state side or exploitative merely on the NAG side. Ideational categories are used to detail the active and passive support, including: 1. safe haven to NAG members, 2. safe haven to NAG members, 3. headquarters, 4. training camp, 5. training, 6. weapon and logistic aid, 7.financial aid, 8. transport of military equipment, 9. troop (only for active support), 10. other kind of support. \\

\begin{figure}[b]
\hskip -0.4cm
\includegraphics[width=1.0\textwidth]{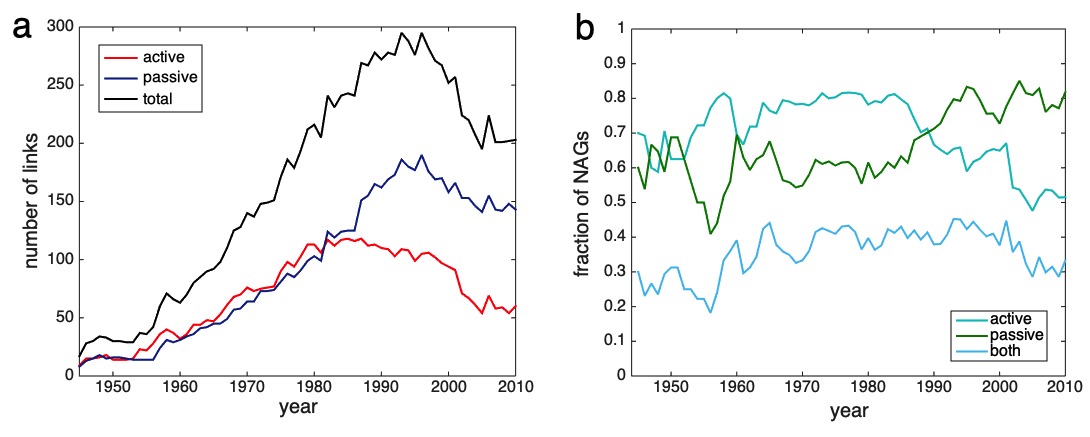}
\caption{\label{fig:NUM_NODE_LINK} Active and passive supporting relations in NAG-HS network. \textbf{a,} Number of links (note that a link can contain active and passive support simultaneously). \textbf{b,} Fraction of NAGs that gain active, passive and both types of support.
}
\end{figure}

\begin{figure}[h]
\vskip 2cm
\hskip -0.2cm
\includegraphics[width=1\textwidth]{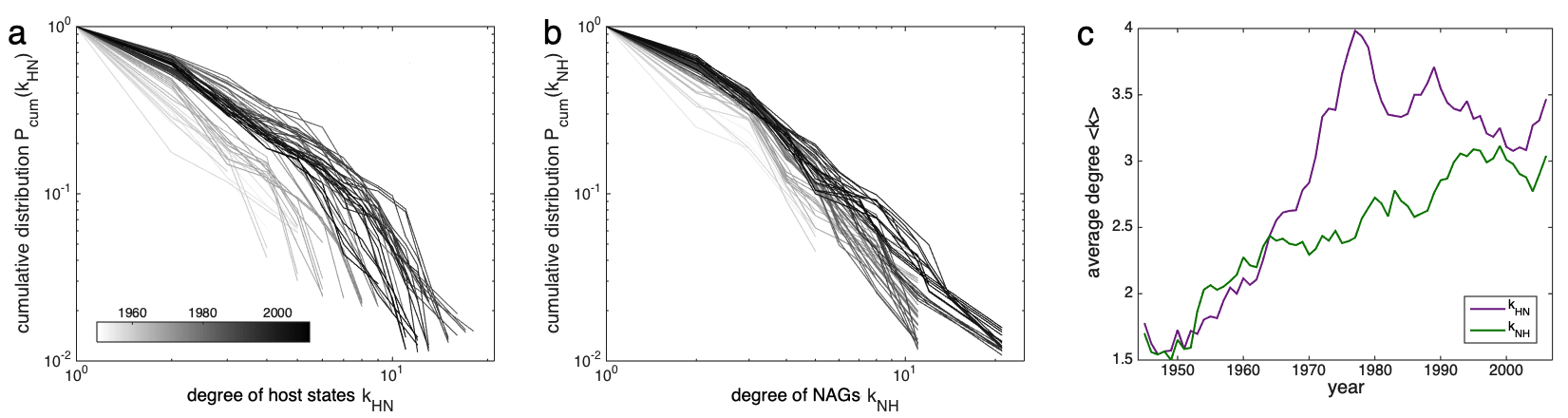}
\caption{\label{NAG_KHN_KNH_AGG_CUM_abc}Cumulative degree distribution of temporal networks for (a) HSs and (b) NAGs, with the average degree shown in panel (c). The degrees are measured based on aggregated networks in a 5-year time window. }
\end{figure}

\noindent\textbf{Basic statistics.} Generally, the NAG-HS supporting network experienced an almost monotonic increase and then decrease in its size, in terms of either number of nodes or links, reaching its maximum in the early 90s, corresponding to several years after the end of the cold war (ECW, around 1990), as shown in Fig. 1a in the main text and Fig. \ref{fig:NUM_NODE_LINK}a. The fractions of links that correspond to active and passive support have also changed drastically from the 1980s on (see inset of Fig. 1b in the main text). Along with this, the average number of attached counterpart actors has also changed: While the average number of supporting HSs per NAGs increased from around 1.5 to 3 for all years, that of attached NAGs per HS increases from 1.5 to 4 in 1990 (right at ECW) and decreases thereafter (Fig. \ref{NAG_KHN_KNH_AGG_CUM_abc}c). The latter suggests that the ECW has a significant influence on several aspects of the supporting network, alongside the nestedness and modularity discussed in the main text.

\begin{figure}[t]
\centering
\includegraphics[width=1\textwidth]{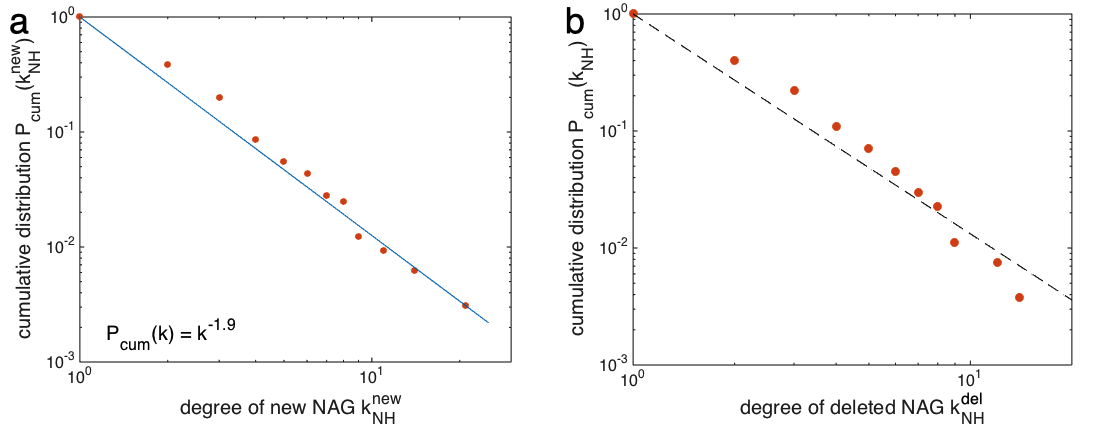}
\caption{\label{NAG_DEG_NEW_DEL_NAG}Cumulative degree distributions of (a) newly added and (b) deleted NAGs.}
\end{figure}

The degree distribution remains similar in all time slices, as shown in Fig. \ref{NAG_KHN_KNH_AGG_CUM_abc}a and \ref{NAG_KHN_KNH_AGG_CUM_abc}b (except those of the HSs in the early years). Moreover, both degree distributions of newly added NAGs and deleted NAGs follow approximately a power law (whereas either newly added and deleted states only have degree 1, 2 or very seldom 3) (Fig. \ref{NAG_DEG_NEW_DEL_NAG}).\\ 

\noindent\textbf{Assembly and disassembly.} The tendencies of NAGs in attaching to and detaching from high fitness (degree) HSs have been demonstrated in the main text. Alternatively, we can observe how NAGs acquire and remove HS supporters. The histograms shown in Fig. \ref{att_detach}a and d are from the standpoint of NAGs. It is notable that a significant fraction of HSs attach to NAGs that are newly joining the supporting network (corresponding to the zero-degree bar in Fig. \ref{att_detach}a). Turning to relative probabilities (see Fig. \ref{att_detach}b and e), we again observe that NAGs with a higher degree is more likely to acquire support from newly joining HSs ($T^+(k_{NH})\sim k^{0.93}_{NH}$) and also lose support from outgoing (no longer support any NAGs) HSs ($T^-(k_{NH})\sim k^{0.78}_{NH}$). The sublinear dependence in both processes again suggests generalized (sublinear) preferential attachment (PA) and detachment (PD). Due to the probable scarcity of data points, an obscure yet generally positive dependence (see $R^-(k_{NH})$ in Fig. \ref{att_detach}f) can be seen by NAGs losing connections from incumbent HSs (remaining in the network). The relative probability $R^-(k_{NH})$ is around 1 for all degrees, as shown in Fig. \ref{att_detach}c, which suggests otherwise that the outgoing of HSs from the network occurs randomly at NAGs with almost no bias.\\

\begin{figure}
\hskip -0.2cm
\includegraphics[width=1\textwidth]{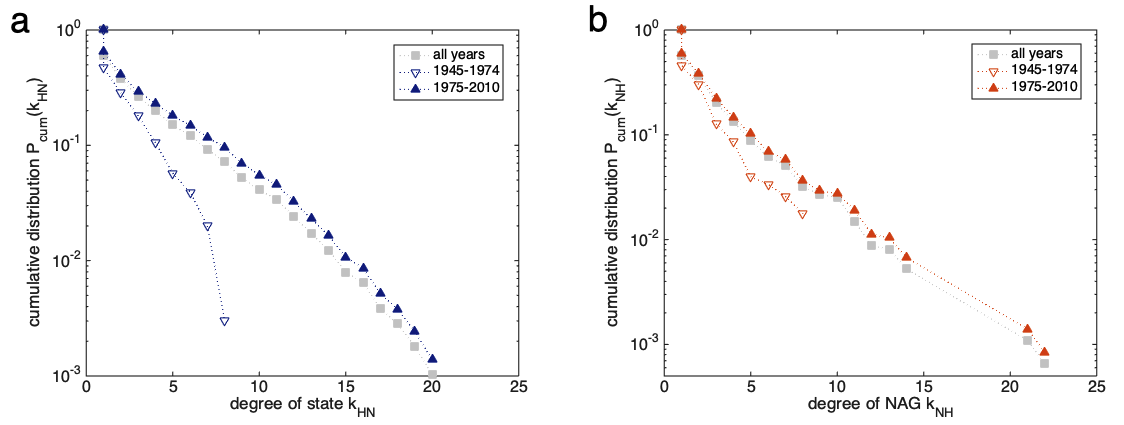}
\caption{\label{deg_PI_PII} Cumulative degree distributions of (a) hosting states and (b) NAGs for the periods 1945-1974, 1975-2010 and all years. The results were calculated by counting nodes' degrees of all 5-year aggregated networks in the respective time periods.}
\end{figure}

\begin{figure}
\hskip -0.4cm
\includegraphics[width=1.05\textwidth]{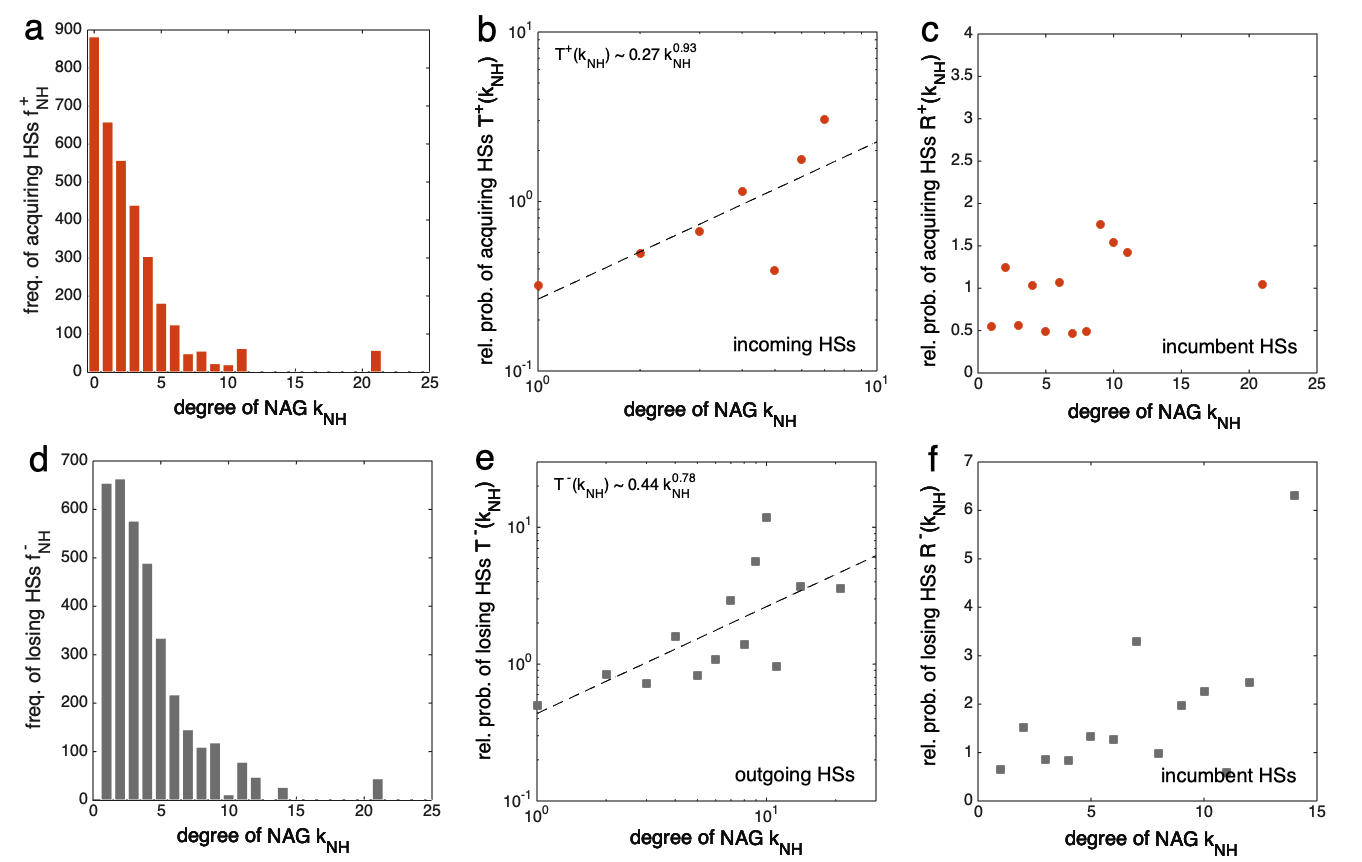}
\caption{\label{att_detach} Attachment and detachment of host states. \textbf{a, d}, Histograms for the frequencies of acquiring and losing links from HSs by NAGs with degree $k_{NH}$ (zero-degree for new NAGs). \textbf{b}, Relative probability $T^+(k_{NH})$ that an incumbent NAG with $k_{NH}$ previous connections acquires links from newly joining HSs. \textbf{c}, Relative probability $R^+(k_{NH})$ that an incumbent NAG acquires new links from incumbent HSs. \textbf{e}, Relative probability $T^-(k_{NH})$ that an incumbent NAG loses links from HSs that depart the network. \textbf{g}, Relative probability $R^-(k_{NH})$ that an incumbent NAG loses links from incumbent HSs.}
\end{figure}

\noindent\textbf{Change in nestedness.} The difference in the nestedness before and after 1975, compared with the null model (shown in Fig. 3a in the main text), is mainly due to the changed broadness of degree distribution, which is known to be positively correlated with the nestedness \cite{Jonhson:2013S}. We compare in Fig. \ref{deg_PI_PII} the cumulative degree distribution in the two periods, respectively, 
which has been broadened due to the preferential attachment that dominates over the detachment processes. Yet, as discussed in the main text, the end of Vietnam War around 1975 may have accelerated the broadening due to the emergence of several hubs of giant powers, which might have resorted more often to proxy wars and become more willing to support NAGs across different parts of the world (HS-hubs). Such support contributed to the endurance and level of violence of many armed groups, some of which then became the big NAG-hubs. The rise of both HS- and NAG-hubs rendered higher overlaps of partnerships, which contributes to the nestedness. \\

\begin{figure}[h]
\hskip 0.9cm
\includegraphics[width=0.8\textwidth]{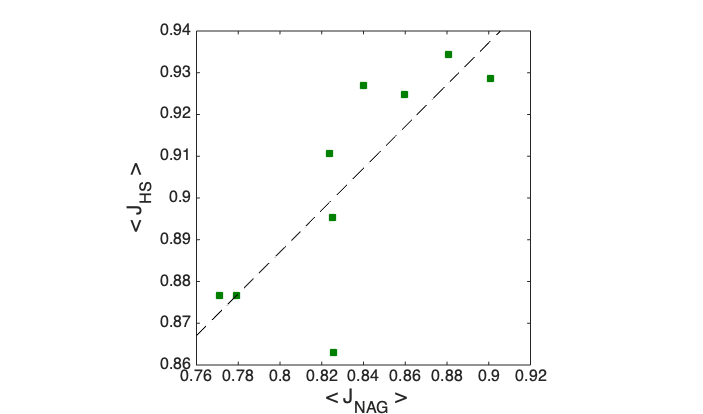}
\caption{\label{fig:JJ_avg_corr}Levels of fluctuation in submodules. Positive correlation between the averages of the Jaccard indices of NAG- and HS-submodules of the 9 major modules ($P = 0.01$, two-sided t-test). 
}
\end{figure}

\begin{figure}[h]
\hskip 0.2cm
\includegraphics[width=1\textwidth]{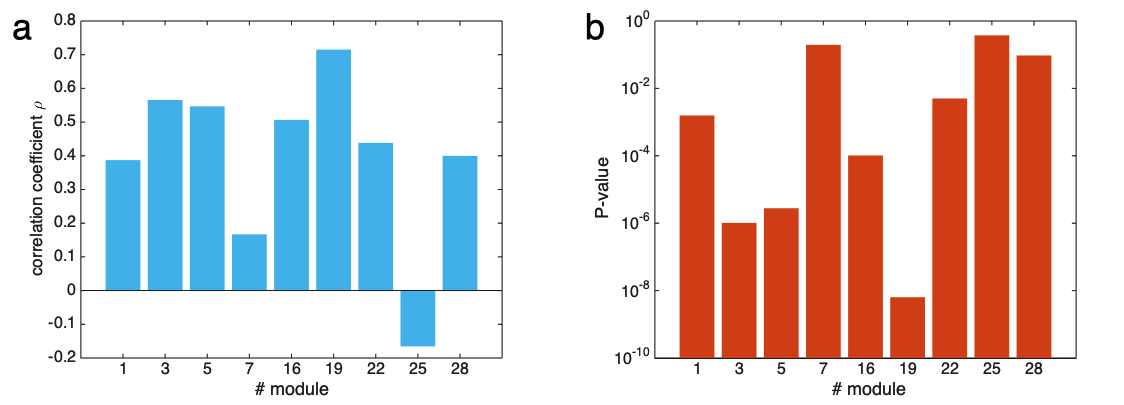}
\caption{\label{fig:JJ_corr_coef_pv}Temporal correlation of NAG- and HS-submodules of major modules, measured by \textbf{(a)} Pearson coefficients and \textbf{(b)} the P-values (two-sided t-tests). 
}
\end{figure}

\noindent\textbf{Jaccard indices.} We calculate the temporal Jaccard indices to track the 
membership of the submodules of the 9 major modules, which all turn out to be high (close to 1). The NAG- and HS-submodules show similar levels of fluctuation over the time span of their existence, as indicated by the correlation of the averages of the Jaccard indices of the corresponding submodules (see Fig. \ref{fig:JJ_avg_corr}). In individual modules, for 7 of the 9 major modules, the temporal NAG- and HS-submodules are highly correlated ($P < 0.1$, two-sided t-test), as indicated by their Pearson coefficients shown in Fig. \ref{fig:JJ_corr_coef_pv}. 

\newpage

\newpage

\clearpage
\newpage


\clearpage
\newpage
\section*{S2. Robust Partition of Temporal Network}

We first apply the temporal community detection algorithm, GenLouvain, to partition the NAG-HS supporting network \cite{Mucha:2010S}. The algorithm, based on the concept of Markov stability, aims to maximize a single objective function by finding a systematic optimal partition of the network throughout all time slices (years). 
\begin{eqnarray}
\label{quality}
Q_M=\frac 1 {2\mu} \sum\limits_{ij\alpha\beta}\left[\left(A_{ij\alpha}-\gamma\frac{k_{i\alpha}k_{j\alpha}}{2m_{\alpha}}\right)\delta_{\alpha\beta}+\delta_{ij}C_{j\alpha\beta}\right]\delta(\sigma_{i\alpha},\sigma_{j\beta})
\end{eqnarray}
where $A_{ij\alpha}$ is the adjacent matrix at time $\alpha$, $\gamma$ is the resolution parameter, $k_{i\alpha}$ is the degree of node $i$ in the $\alpha$'s time slice, $m_{\alpha}=\sum\limits_i k_{i\alpha}$. Distinct from any static detection algorithm, where slices are isolated, an interslice coupling strength $C_{j\alpha\beta}$ is used, which equals 0 or $\omega$ indicating the absence or presence of an interslice link at node $j$. The partition of the $\alpha$th slice is denoted by $\sigma_{i\alpha}$, and $\delta(\sigma_{i\alpha},\sigma_{j\beta})$ equals 1 when the nodes $i$ at time $\alpha$ and $j$ at time $\beta$ are partitioned into the same module, and 0 otherwise. In general, one may adjust the intraslice resolution $\gamma_{\alpha}$ and interslice resolution $\omega$. However, we have found no specific values that correspond to characteristic resolutions and hence chosen default values for these two resolutions, i.e., $\gamma=1$ and $\omega=1$.  

\begin{figure}[b]
\centering
\includegraphics[width=1\textwidth]{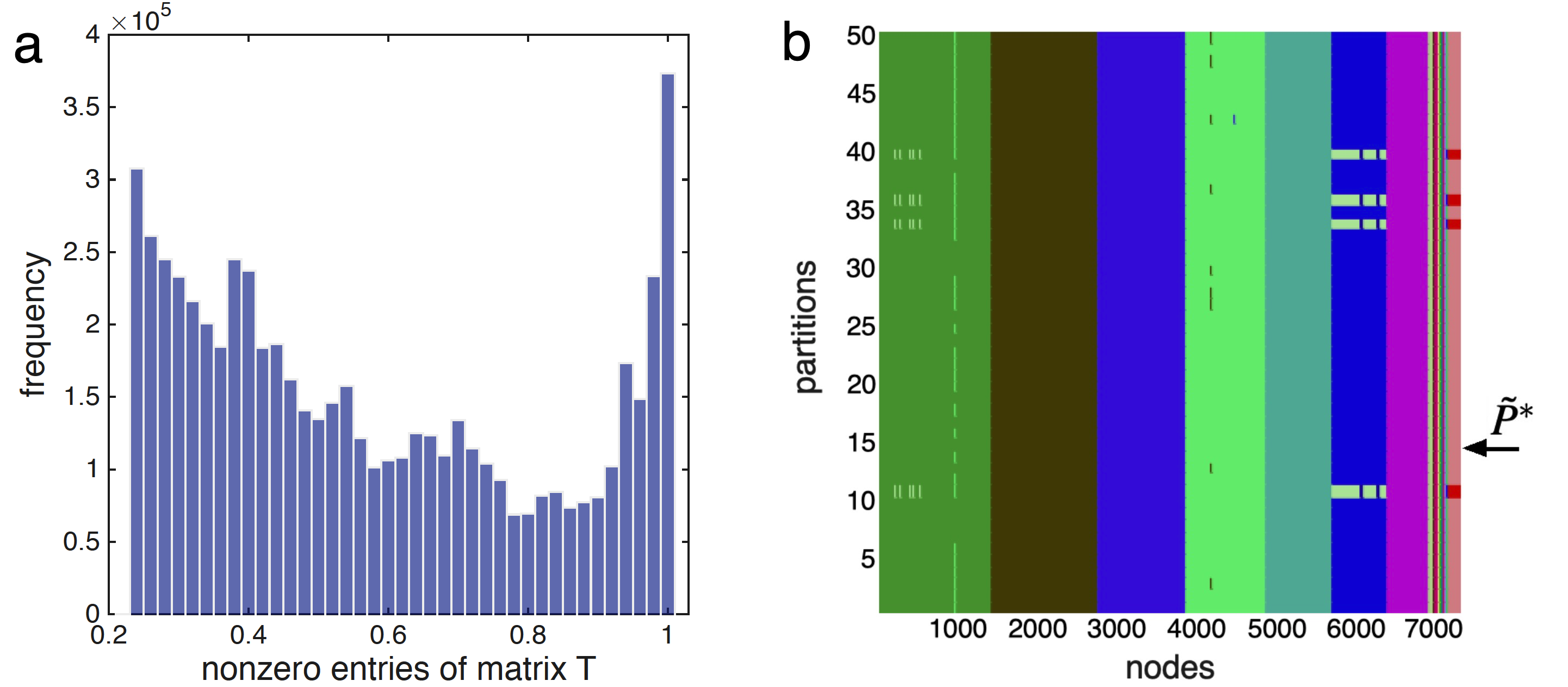}
\caption{\label{Tmat} Representative partition. \textbf{a,} Histogram of nonzero entries of the association matrix. Each entry corresponds to the relative frequency of a pair of nodes belonging to the same module, according to the partitioning of the original temporal adjacency matrices. A threshold is set to discount the incidences from random partitioning. \textbf{b,} Partitions of the association matrix. 50 partitions are realized (each colour corresponds to a module) and turn out to be almost identical. The arrow points to the representative partition $\tilde{P}^*$, according to maximizing the AMI with other partitions.}
\end{figure}

By using this definition, one can then use any heuristic algorithm to maximize $Q$, and in particular, we have used Louvain's algorithm for this purpose, which is implemented as the GenLouvain algorithm. Note that here we use the algorithm for the bipartite network, and thus the entries in the adjacency matrix $A_{ij}$ are nonzeros when $i$ and $j$ correspond to the NAG and HS parties, respectively, that is, $i=1\dots S_{NAG}$ and $j=(S_{NAG}+1)\dots (S_{NAG}+S_{HS})$.

Note that by using any Louvain-like algorithm, there can still be some arbitrariness in partitioning the same network, denoted as $\{P_s\}$ (here we have obtained 50 partitions). The specific $s$th partition is denoted by $P_s(t)=\cup_i M_{s,i}(t)$, where the subscript $i$ denotes the $i$th module and $t$th year. We can then find a representative partition by using a stabilizing technique proposed by Bassett, et al \cite{Bassett:2013S}, if the modular structure is truly nontrivial. 

We construct a nodal association matrix $T$, each entry of which is the frequency of a pair of actors being partitioned in the same module according to the previously obtained partitions $\{P_s\}$. Note that this association matrix $T$ is consisted of all NAGs and HSs over the entire timespan as the column and row indices, respectively. To eliminate the cases that two nodes are partitioned into one module purely by chance, we set any element $T_{ij}$ in the original association matrix $T$ that exceeds the maximum entry of a null association matrix $T_r$ to 0, so that only statistically significant relationships between nodes are retained. The null matrix $T_r$ is constructed by reassigning nodes uniformly at random to $N_m$ communities of mean size $S_r$, where $N_m$ is the number of modules obtained previously. 

By partitioning the matrix $T$ with the GenLouvain algorithm again (50 times), we find that all partitions $\{\tilde{P}_s\}$ of $T$ are almost identical, indicating the existence of a highly robust modular structure. The remaining arbitrariness has thus become very small at this step. A representative partition $\tilde{P}^*$ can then be selected, which should have a maximal average Adjusted Mutual Information (AMI) with all other partitions $\{\tilde{P}_s\}$. The AMI is defined as follows \cite{Vinh:2010S}
\begin{eqnarray}
\label{ami}
AMI(\tilde{P}_1,\tilde{P}_2)=\frac{MI(\tilde{P}_1,\tilde{P}_2)-E\{MI(\tilde{P}_1,\tilde{P}_2)\}}{\max\{H(\tilde{P}_1),H(\tilde{P}_2)\}-E\{MI(\tilde{P}_1,\tilde{P}_2)\}}
\end{eqnarray}
which is a variation of the mutual information $MI$ of $\tilde{P}_1$ and $\tilde{P}_2$. Here, $H$ is the entropy of the partition and $E$ is the expected value. 

\begin{figure}
\hskip -0.0cm
\includegraphics[width=1.0\textwidth]{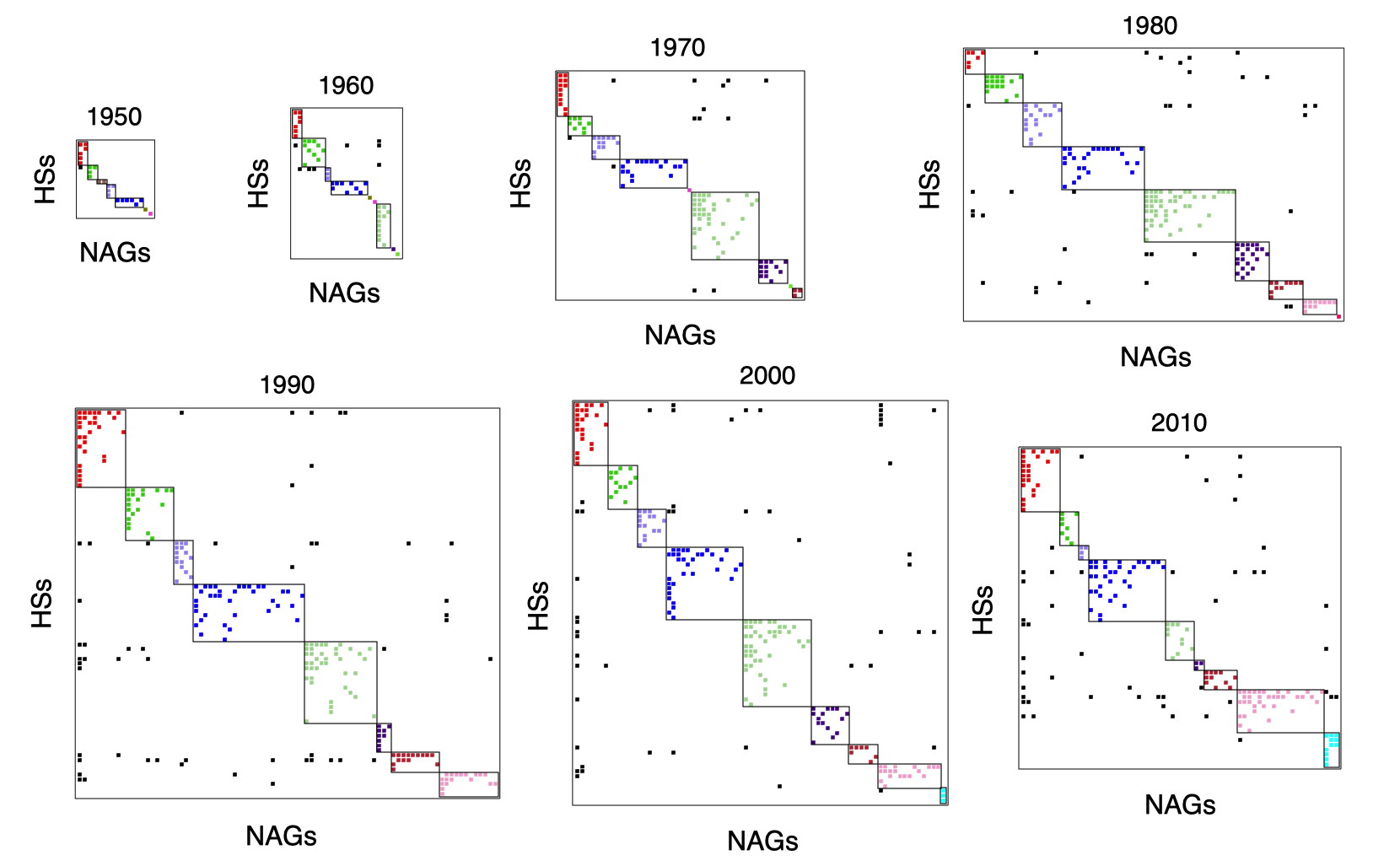}
\caption{\label{Mod_State_Nag}Modular structure for the NAG-HS network for every decade from year 1950 to 2010. The adjacency matrices are sorted according the module-membership and each module is assigned a unique colour, which does not change over time.}
\end{figure}

\begin{figure}
\hskip -0.2cm
\includegraphics[width=1.0\textwidth]{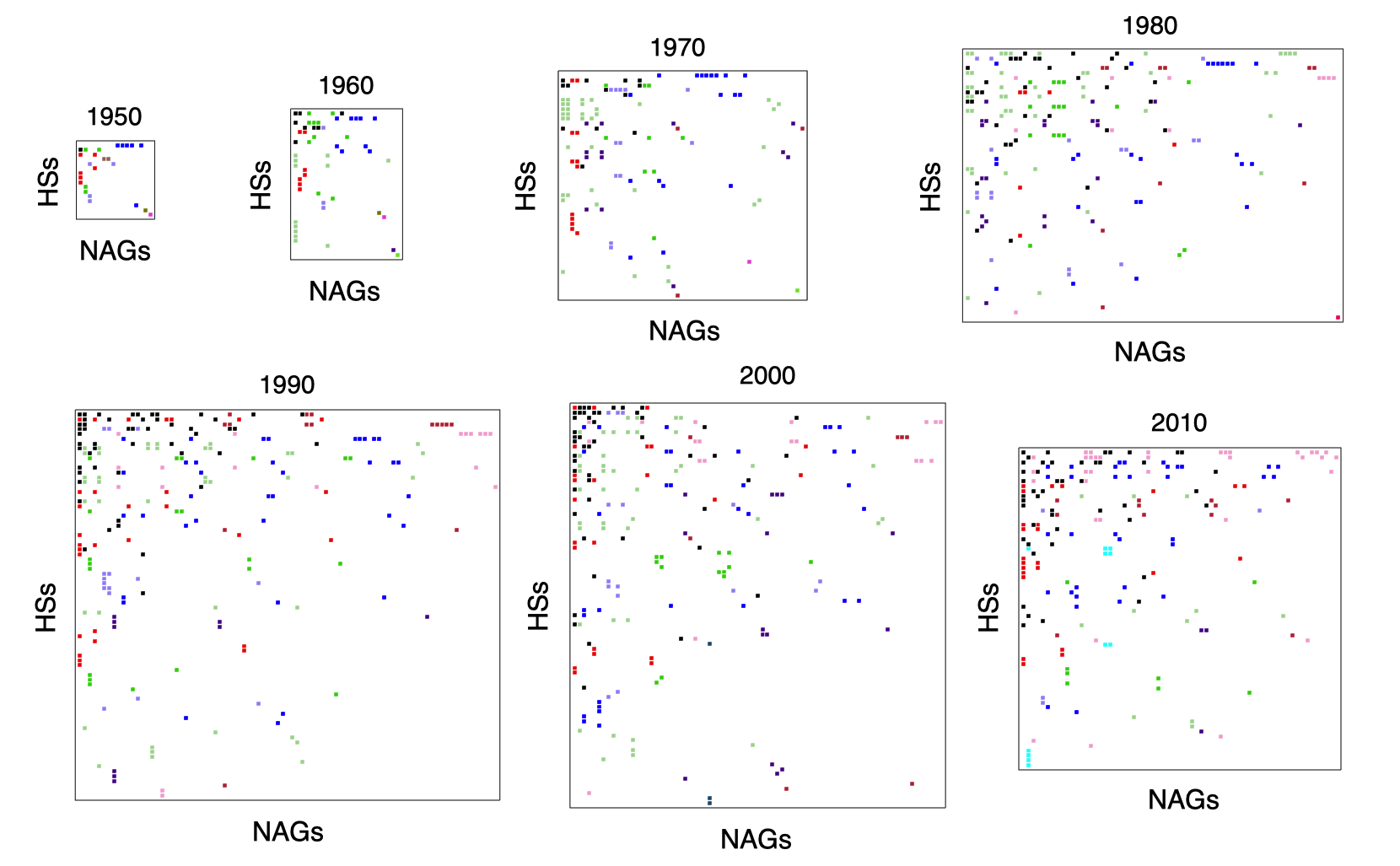}
\caption{\label{Nest_State_Nag}Nested structure for the NAG-HS network for every decade from year 1950 to 2010. The colour represents the module-membership, identical to that in Fig. \ref{Mod_State_Nag}.}
\end{figure}

The rows and columns of the adjacency matrices are sorted to show the resulted bipartite modular structures, as demonstrated for the time slices every 10 years in Fig. \ref{Mod_State_Nag}, where the same colour is used for the same module over all years. As shown in Tab. \ref{MM1970} and \ref{MM2000}, the membership in all modules has changed significantly from year 1970 to 2000. Only 7 NAGs in 1970 reappeared in 2000, as marked in bold. Also, note that not all the 9 major modules appear in each of these network snapshots and in some years, some of the major modules may be very small. For example, only 7 major modules appear in 1970, as shown in Tab. \ref{MM1970}. It is exactly the merit of the temporal detection algorithm that the continuity of the same module is maintained even if some of the modules are insignificant in several time slices \cite{Fortunato:2010S}. The modules have exhibited apparent geographical similarity, yet distant nations can also be involved in supporting certain regional NAGs in a time period. Meanwhile, the adjacency matrices are alternatively sorted by the degree (generalists or specialists), which show the nested structures in Fig. \ref{Nest_State_Nag}. 

\begin{figure}
\hskip -0.4cm
\includegraphics[width=1\textwidth]{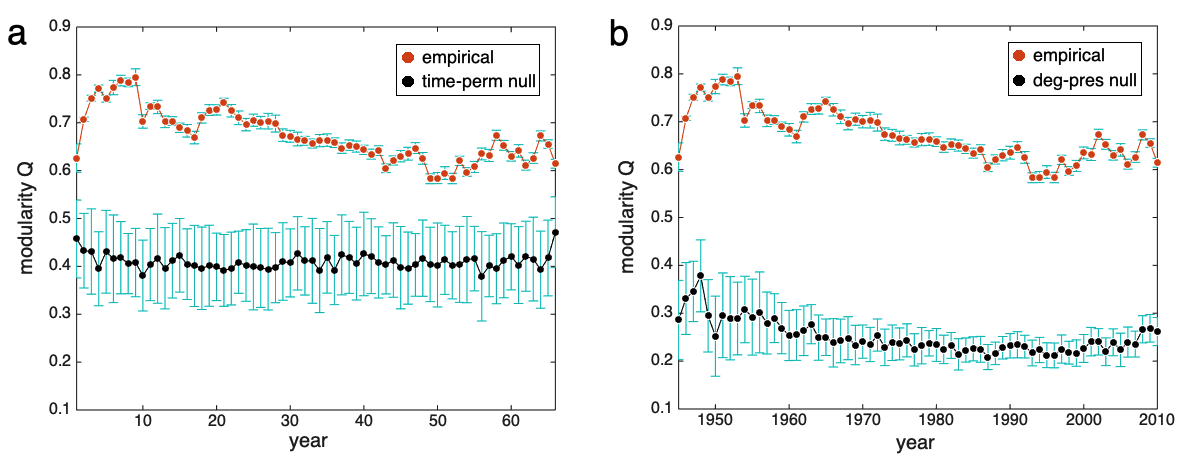}
\caption{\label{NG_MOD_YR_Perm_Deg_Null}Different null models for the detection of modularity: (a) Permutation of time slices; (b) Randomly shuffled within-slice links by preserving the same degree sequence.}
\end{figure}

We have used three null models, which are constructed as follows \cite{Bassett:2011S}: 1.the probability of each entry of the adjacency matrix being occupied is the average of the occupation probabilities of the row and column (Fig. 3c); 2. randomly permuted year slices (Fig. 3c and Fig. \ref{NG_MOD_YR_Perm_Deg_Null}a); and 3. randomly shuffled within-slice links by keeping the same degree sequence (Fig. \ref{NG_MOD_YR_Perm_Deg_Null}b).
















\clearpage
\newpage

\clearpage
\newpage

\clearpage
\newpage

\clearpage
\newpage
\setlength{\arrayrulewidth}{0.25mm}
\begin{center}
\captionof{table}{Module Members in 1970} \label{MM1970} 
\begin{tabular}{ |p{0.1\textwidth}|p{0.42\textwidth}|p{0.42\textwidth}|} 
\hline
 \bf Module ID & \bf Non-State Armed Groups & \bf Host States\\
 \hline 
 M1 & KDP, Anya Nya & Congo, Uganda, Kenya, Ethiopia, Iran, Turkey, Israel, USA, UK\\
 \hline 
 M2 & FNL, MPLA, \textbf{FARC}, FSLN, Pathet Lao & Cuba, Costa Rica, Soviet Union, North Vietnam\\
 \hline 
 M3 & CPB, CPT, NNC, FNLA, \textbf{UNITA}, FNLA & Zaire, South Africa, China, Cambodia, Laos\\
 \hline 
 M4 & \textbf{KNU}, NMSP, CPM, KNPP/KA, SSA, SURA, MNF, KR, MIM, Jam'iyyat-i Islami-yi Afghanistan, \textbf{UNLF}, \textbf{Patani insurgents}, PBCP, CPM & India, Pakistan, Burma, Thailand, Malaysia, Germany\\
 \hline 
 M5 & KDPI, Royalists, \textbf{PFLP}, PFLP-GC, \textbf{Fatah}, Muslim Brotherhood, PIRA, PFLO, ETA, ELF, Frolinat, Republic of Biafra, Sudanese Communist Party, BLF & Ivory Coast, Algeria, Libya, Iraq, Egypt, Syria, Lebanon, Jordan, Saudi Arabia, South Yemen, Bahrain, Ireland, France, Germany\\
 \hline 
 M6 & OPM, Frelimo, SWAPO, ZAPU, ZANU, ANC & Tanzania, Angola, Zambia, Botswana, Sweden\\
 \hline 
 M7 & EPLF, Jamaat al-Muslimeen & Sudan, Canada\\
 \hline
 MX & Tibet, ANLP & Nepal, Bangladesh\\
 \hline
\end{tabular}
\end{center}

\setlength{\arrayrulewidth}{0.25mm}
\begin{center}
\vskip 2cm
\captionof{table}{Module Members in 2000} \label{MM2000} 
\begin{tabular}{ |p{0.1\textwidth}|p{0.42\textwidth}|p{0.42\textwidth}| } 
\hline
 \bf Module ID & \bf Non-State Armed Groups & \bf Host States\\
 \hline 
 M1 & Hamas, Junbish-i Milli-yi Islami, MEK, ETA, UCK, MDJT, MFDC & Albania, Bosnia and Herzegovina, Guinea-Bissau, Gambia, Turkey, Yemen, Kuwait, Qatar, UAE, UK, France, Germany, Italy\\ 
 \hline 
 M2 & PIRA, RIRA, \textbf{FARC}, ELN, EPL, Sendero Luminoso & Cuba, Panama, Columbia, Venezuela, Ecuador, Bolivia, Ireland, Croatia, North Vietnam\\ 
 \hline 
 M3 & MQM, SPLM/A, LURD, FLEC-R, Cocoyes, \textbf{UNITA} & Ivory Coast, Guinea, Sierra Leone, South Africa, Namibia, Israel, USA, Portugal\\ 
 \hline 
 M4 & \textbf{KNU}, God's army, NSCN-K, RCSS/SSA-S, CPN-M, NLFT, PLA, \textbf{UNLF}, PREPAK, LTTE, ULFA, GAM, NDFB, \textbf{Patani insurgents}, PBCP, MNDAA & North Korea, India, Bhutan, Bangladesh, Burma, Thailand, Cambodia, Malaysia, Singapore, Indonesia, China, Australia, Sweden, Norway, Denmark\\
 \hline 
 M5 & \textbf{PFLP}, \textbf{Fatah}, PIJ, Hizb-i Wahdat, UIFSA, PKK, Devrimci Sol, Republic of Nagorno-Karbakh, Republic of South Ossetia, Hezbollah, RFDG, MPA, RUF, AFRC & Ukraine, Armenia, Burkina Faso, Liberia, Ghana, Algeria, Libya, Iran, Iraq, Egypt, Syria, Lebanon, Jordan, South Yemen, Greece, Russia, Netherlands, Germany\\
 \hline 
 M6 & MLC, RCD, Palipehutu-FNL, Frolina, CNDD-FDD, FLEC-FAC, Ninjas, ALiR & Congo, Zaire, Uganda, Tanzania, Burundi, Rwanda, Angola, Zimbabwe\\
 \hline 
 M7 & LRA, UNRF II, EIJM-AS, ONLF, OLF, Jamaat al-Muslimeen & Kenya, Eritrea, Sudan, Canada\\
 \hline 
 M8 & PWG, MCC, NSCN-IM, ASG, Hizb-i Islami-yi Afghanistan, Taleban, Kashmir Insurgents, TNSM, IMU, GIA, al-Qaida, ETIM, IMU & Saudi Arabia, Afghanistan, Tajikistan, Pakistan, Belgium\\
 \hline 
 M9 & AQIM & Spain, Mali, Senegal\\
 \hline
\end{tabular}
\end{center}

\end{document}